\documentclass[journal]{IEEEtran}
\usepackage[utf8x]{inputenc}
\usepackage{amsmath}
\usepackage{amssymb}
\usepackage{latexsym}
\usepackage{gensymb}
\usepackage{array}
\usepackage{adjustbox}
\usepackage{anyfontsize}
\usepackage{lipsum}
\usepackage{float}
\restylefloat{table}
\usepackage{ifpdf}
\usepackage{graphicx}
\usepackage{mathrsfs}

\ifpdf
\DeclareGraphicsExtensions{.pdf, .jpg, .tif, .png}
\else
\DeclareGraphicsExtensions{.eps, .jpg}
\fi

\usepackage[acronym,nonumberlist,numberedsection=autolabel,
			section,toc]{glossaries}

\usepackage[square, numbers, sort&compress]{natbib}

\ifCLASSINFOpdf
\else
\fi
\begin{document}
%
\title{A Review of Contemporary Atomic Frequency Standards}
%
%
%

\author{Bonnie~L.~Schmittberger
        and~David~R.~Scherer
\IEEEcompsocitemizethanks{B. Schmittberger is with The MITRE Corporation, McLean, VA. D. Scherer is with The MITRE Corporation, Bedford, MA. e-mail: dscherer@mitre.org}
}

\maketitle

\begin{abstract}
Atomic frequency standards are used to generate accurate and precise time and frequency, enabling many communications, synchronization, and navigation systems in modern life. GPS and other satellite navigation systems, voice and data telecommunications, and timestamping of financial transactions all rely on precise time and frequency enabled by atomic frequency standards. 

This review provides a snapshot and outlook of contemporary atomic frequency standards and the applications they enable. We provide a concise summary of the performance and physics of operation of current and future atomic frequency standards. Additionally, examples of emerging frequency standard technologies and prototype demonstrations are presented, with a focus on technologies expected to provide commercial or military utility within the next decade. 

We include a comparison of performance vs. size and power for current atomic frequency standards, and we compare early prototypes of next-generation frequency standards to current product trends. An empirical relationship between frequency standard performance and product size is developed and discussed. Finally, we provide a mapping between applications and frequency standard technologies.
\end{abstract}


%
\IEEEpeerreviewmaketitle

\section{Introduction}
%
%
%
%

\IEEEPARstart{A}{tomic} frequency standards enable enhanced autonomy, resiliency, and integrity for critical navigation, communications, sensing, and other applications. Commercial frequency standards that have emerged within the past ten years, such as the chip-scale atomic clock (CSAC), represent extraordinary advances in engineering and have enabled the expansion of autonomous timekeeping to a wider range of mobile platforms. Next-generation atomic frequency standards could further reduce reliance on  Global Navigation Satellite Systems (GNSS), enable novel tests of fundamental physics which require extreme levels of precision, and are expected to change the definition of the second.

Here we present a survey of current and future atomic frequency standards. We analyze how frequency standard performance scales with physical parameters, project future frequency standard performance, and identify key applications where we anticipate these frequency standards will make a substantial impact. Time and frequency terminology and an overview of atomic frequency standard operation are presented in Sections~\ref{terminology} and \ref{overview}. Section~\ref{current} includes a description current atomic frequency standards, including physics of operation and performance characteristics. This section includes a list of commercial products that can be purchased in volume at the time of writing, as well as a few recent product introductions and preliminary data sheets, based on current publicly available information. Emerging technologies such as research and development prototypes developed by universities, national laboratories, and private industry are described in Section~\ref{next_generation}. Quartz and other types of non-atomic oscillators have been omitted.

In order to provide tangible comparisons, specific product names, vendors, prototypes, and research programs are included. The reader is cautioned, however, that product specifications can change over time, and that data sheets may not be indicative of actual performance.  Regarding emerging technologies, we emphasize those that appear to have the most promise for development as commercial products or useful prototypes within the next decade.

Reviews of atomic frequency standards can be found in several articles \cite{hellwig_review, lewis1991introduction, beehler_review, audoin_review, mccoubrey_review} and books \cite{vanier1989quantum, riehle2006frequency}. Additionally, several reviews have highlighted recent progress in specific areas, such as optical frequency standards \cite{ludlow_optical}, frequency standards based on coherent population trapping \cite{vanier_cpt}, atomic fountains \cite{wynands_fountain}, and space frequency standards \cite{bhaskar1996historical, mallette2007historical}. This review provides a modern update to the literature, and differs from previous reviews in the quantity of information on current commercially available frequency standards and the introduction of novel comparison charts quantifying frequency standard performance.

\subsection{Time and Frequency Terminology} \label{terminology}
Standard terminology used in the field of time and frequency is defined in several sources in the literature \cite{lombardi2005fundamentals, allan1987time}. In this section, we summarize the most important terms relevant to atomic frequency standards.

An oscillator, or frequency standard, is a device that produces a periodic signal. An atomic frequency standard is a frequency standard whose basic resonant system is an atom or molecule experiencing a transition between two quantized energy levels. An atomic frequency standard provides a stable frequency output, typically at 10~MHz. An atomic clock is a continuously operating atomic frequency standard coupled with a counter and initialization time, providing the user with time as well as frequency. Commercial frequency standard products typically have two outputs: one is an oscillatory signal at 10~MHz, the other is a square-wave one pulse-per-second (1 PPS) output. When powered on, these products operate as frequency standards, and the 10~MHz output is typically used for applications that require a stable frequency. The same device can be operated as an atomic clock by counting cycles since an initialization time on a counter. For applications that require time, the 1 PPS output is commonly used.

Frequency is the rate at which repetitive phenomena occur over time. Current atomic frequency standards generate an output in the radio frequency (RF) region of the electromagnetic spectrum. The dimensionless, normalized frequency $y(t)$ is related to the instantaneous frequency $f(t)$ and the nominal frequency $f_0$ by
\begin{equation} 
\label{frac_freq}
y(t) = \frac{f(t)-f_0}{f_0}.
\end{equation}
In Eq.~\ref{frac_freq}, $f_0$ is typically 10~MHz, so if the instantaneous frequency is off by 1~mHz, the fractional frequency (or fractional frequency offset) $y$ is $1 \times 10^{-10}$.

Instability is a characteristic of an oscillator that describes how well it produces the same frequency over a given time interval \cite{lombardi2005fundamentals}. In the time and frequency literature, this characteristic is referred to as instability, fractional frequency instability, or fractional frequency stability. The terms instability and stability are often used interchangeably, with instability favored by the high-performance frequency standard community and stability favored by the commercial vendor base. Instability is typically characterized by the Allan Deviation (ADEV), denoted by $\sigma_y(\tau)$, equal to the square root of the Allan Variance (AVAR), denoted by $\sigma_y^2(\tau)$ \cite{allan1987time}, where $\tau$ is the interval time. For the characterization of frequency standards and oscillators, the Allan variance, rather than a standard variance, is used because the noise behavior displayed by frequency standards is non-stationary. A detailed description of ADEV, along with other variances used in time and frequency metrology, can be found in the literature \cite{NIST_1337}. 

The ADEV $\sigma_y(\tau)$ describes the instability of the fractional frequency $y$ over a given time interval $\tau$. This can be loosely understood as follows: for a frequency standard with a specified instability (ADEV) of $1 \times 10^{-10}$ at 1~second, one can expect a fractional frequency change of $1 \times 10^{-10}$ on average after 1~second of operation; \emph{i.e.} if the frequency standard output frequency is exactly 10~MHz at $t=0$, one can then expect it to have changed by about 1~mHz after 1~second. The flicker floor is the minimum value of ADEV as a function of $\tau$. For longer measurement times the ADEV degrades (increases) due to frequency drift, or aging.

Accuracy refers to the degree of conformity of a measured or calculated value relative to its definition. In the time and frequency literature, accuracy is referenced to the SI definition of the second. The current SI definition of the second is the duration of 9\hspace{0.8mm}192\hspace{0.8mm}631\hspace{0.8mm}770 periods of the radiation corresponding to the transition between the two hyperfine levels of the ground state of the $^{133}$Cs atom (at a temperature of 0~K). In the case of a commercial Cesium Beam Tube, for example the Microchip 5071A, the accuracy specification of $\pm 5 \times 10^{-13}$ describes the range of fractional frequency offset about the nominal 10~MHz output frequency.

High-performance optical frequency standards are instead characterized by the term uncertainty. For example, if two Yb optical lattice frequency standards are compared against each other, one can establish a relative uncertainty, which can approach numerical values of $10^{-18}$~\cite{McGrew2018}. Because the current SI definition of the second is established with reference to a microwave atomic energy level difference in $^{133}$Cs, one cannot define an ``accuracy'' for optical frequency standards. If the current SI definition of the second were defined in terms of the Yb optical transition, an accuracy could be established.

When characterizing instability, accuracy, and uncertainty, lower numbers indicate better performance. Instability refers to how much a frequency standard varies with respect to itself over some time interval, whereas accuracy refers to how much a frequency standard differs from an accepted definition. It is possible for a frequency standard to have good instability and poor accuracy, and vice versa.

Retrace is defined as the fractional frequency offset after cycling an oscillator through a power sequence. While different manufacturers specify this power cycling sequence differently, a common definition is the fractional frequency offset (for example $\pm 2 \times 10^{-11}$) after an on/off/on period of 24 hours/48 hours/24 hours at constant temperature.

The temperature coefficient, or tempco, is defined as the maximum fractional frequency change over a specified temperature, for example $1 \times 10^{-11}$ over 0~$^\circ$C to 70~$^\circ$C. It is not valid to assume that the frequency change with respect to temperature is linear over the specified range.

Time Deviation (TDEV, or $\Delta T$) describes the expected time error of a clock (in units of seconds) after some holdover time and is defined as
\begin{equation}
\label{time_error}
\Delta T(\tau) = T_0 + \frac{\Delta f}{f} \tau + \frac{1}{2} D \tau^2 + \frac{\tau}{\sqrt{3}} M\sigma_y(\tau) + \epsilon (\tau),
\end{equation}
where $\tau$ is the time interval (or holdover time), $T_0$ is the initial time error, $\Delta f/f$ is the initial frequency error, $D$ is the frequency drift (typically quoted in units of fractional frequency drift per month, sometimes called aging), $M\sigma_y(\tau)$ is the modified ADEV, and $\epsilon(\tau)$ describes the integrated environmental errors over time~\cite{riley_handbook}. The modified ADEV is generally equal to the ADEV for $\tau\gtrsim1$~s. In a laboratory environment, a clock's initial time and frequency offsets can be set to a known value with respect to a calibrated standard, and environmental effects can be neglected. In this case, the third and fourth terms in Eq.~\ref{time_error} can be used to calculate the predicted time error of a free-running clock after some holdover time $\tau$. In mobile applications, the environmental effects $\epsilon (\tau)$, such as tempco, magnetic field sensitivity, and vibration sensitivity usually dominate over the other terms, and this expression is of limited utility. Nevertheless, it is used in the comparison charts to follow because it incorporates both the clock's short-term ADEV as well as drift.

Phase noise $\mathscr{L}(f)$ provides a frequency-domain representation of oscillator instability and is expressed as a spectral quantity in dB below the carrier (or nominal output frequency) at a specified frequency offset. For example, a phase noise of -120~dBc/Hz at 10~Hz offset implies that, for a nominal carrier frequency of 10~MHz, the total integrated power within a 1-Hz measurement bandwidth, centered at an offset frequency of 10~Hz from the carrier (i.e. at 10.000\hspace{0.8mm}010~MHz) is 120~dB below the carrier power.

\subsection{Overview of Atomic Frequency Standards} \label{overview}
\begin{figure}[h]
	\centering
	\includegraphics[width=3.5in]{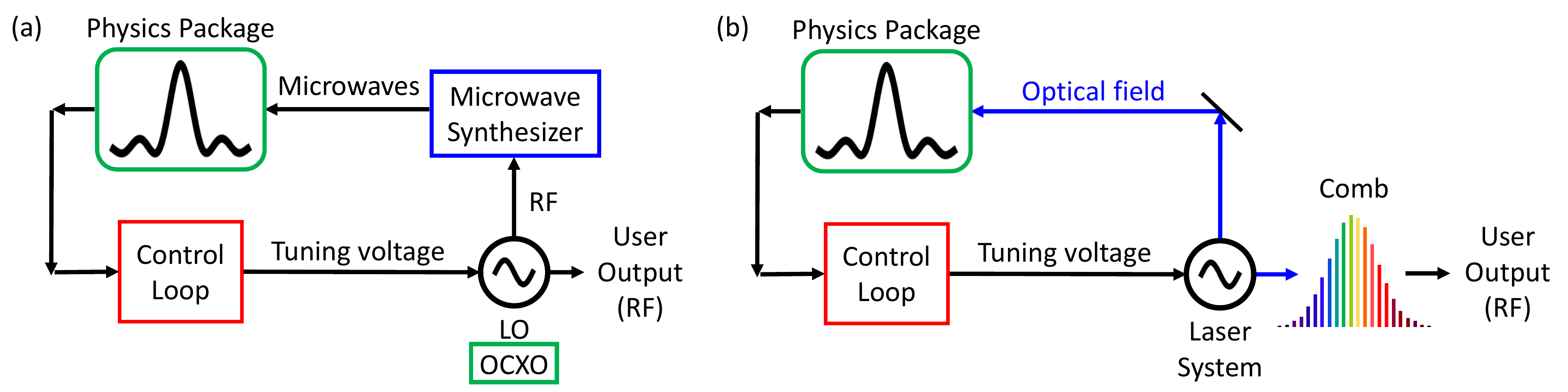} 
	\caption{Schematic of general frequency standard control loop for (a) a microwave frequency standard, and (b) an optical frequency standard. The bandpass filter nature of the atomic resonance response is shown in the physics package. LO = local oscillator, OCXO = oven controlled quartz oscillator.
		\label{clock_control_loop}}  
\end{figure}

An atomic frequency standard consists of a control loop in which a local oscillator (LO) is periodically referenced to an atomic transition frequency. In a microwave frequency standard, the LO is a quartz crystal oscillator. In an optical frequency standard, the LO is a laser. Schematics of microwave and optical frequency standards are shown in Figs.~\ref{clock_control_loop}(a) and (b).

The quartz oscillator has a frequency output in the RF region of the spectrum (typically 10~MHz). A microwave synthesizer  multiplies this RF signal's frequency up to the appropriate microwave frequency that is resonant with the atomic transition. The microwaves are used to interrogate a collection of atoms in the physics package. An error signal is generated based on the difference between the synthesized microwave frequency and the natural atomic resonance frequency. A control loop integrates this error signal and applies a control voltage to steer the LO and ensure that the microwaves are tuned to the atomic resonance. In this way, the long-term stability of the RF output is dictated by the ability of the control loop to adjust the microwaves to atomic resonance, rather than the properties of the LO itself.

Optical clocks work in much the same way, except that the LO is in the optical region of the spectrum (a laser), and an optical frequency comb is required to convert the output signal from the optical domain into the RF domain, as shown in Fig.~\ref{clock_control_loop}(b). These frequency standards are described in more detail in Section~\ref{next_generation}.

There is a time constant, often referred to as the clock `loop tau' or loop attack time, which governs the timescale for regulation of the LO. This timescale is typically on the order of 1~s or less and describes how quickly the control loop can respond to errors. For timescales less than the clock loop tau, the instability is dictated by the short-term instability of the LO. For timescales longer than the clock loop tau, the instability is dictated by the quality of the physics package. The frequency standard instability (ADEV, or $\sigma_y(\tau)$) at time scales longer than the loop tau can be expressed as
\begin{equation}
	\label{q_snr}
	\sigma_y(\tau) = K \frac{1}{Q \times SNR} \tau^{-1/2},
\end{equation}
where $K$ is a unitless constant of order unity and depends on the type of interrogation, $Q$ is the atomic resonance line quality factor (defined as the ratio of the carrier frequency to the linewidth, also unitless), and SNR is the ratio of the signal power to the noise power spectral density. We note that $\sigma_y(\tau)$ is a dimensionless quantity, and SNR in Eq.~\ref{q_snr} has units of $\sqrt{\text{Hz}}$ due to the dependence of the noise on the measurement bandwidth.

\section{Current Atomic Frequency Standards} \label{current}

\subsection{Overview}
\begin{figure}[h]
	\centering
	\includegraphics[width=3.5in]{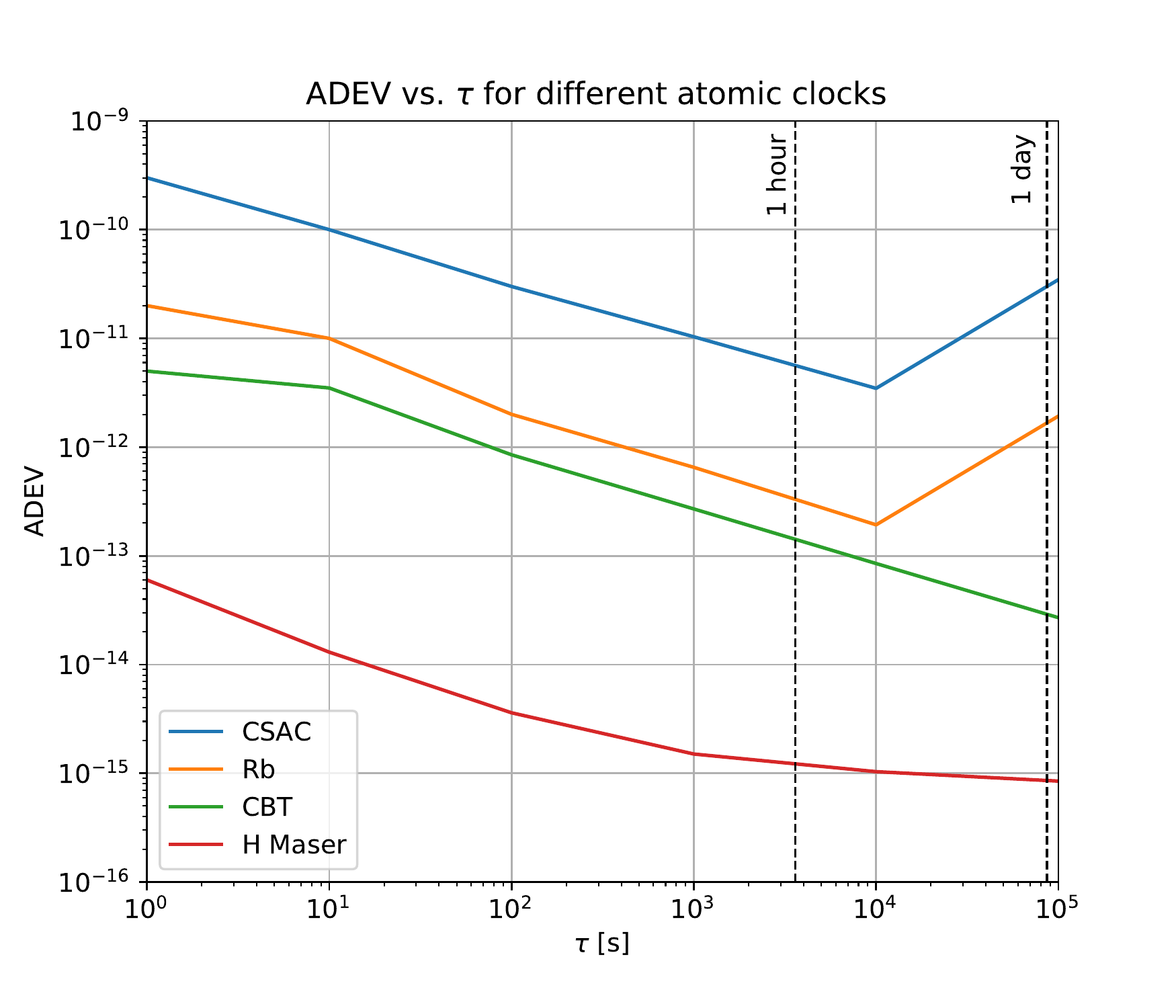} 
	\caption{Comparison of ADEV for different frequency standard technologies. The curves shown from top to bottom are the Microchip CSAC, SRS PRS10 Rb frequency standard, Microchip 5071A Cesium Beam Tube, and Vremya Active Hydrogen Maser. \label{ADEV_summary}}  
\end{figure}

In this section, we describe the physics of operation of current atomic frequency standards and outline their performance in terms of the quantities defined in Section~\ref{terminology}. Current atomic frequency standards can be divided into three main categories. Hydrogen masers are floor-mounted laboratory instruments capable of achieving excellent instability. Cesium Beam Tube (CBT) atomic frequency standards are rack-mounted instruments with excellent accuracy and instability. Gas cell frequency standards range from hand-held to rack-mounted devices and achieve moderate performance. A comparison of ADEV vs. $\tau$ of four different current frequency standard technologies is shown in Fig.~\ref{ADEV_summary}.

\subsection{Hydrogen Maser}

\subsubsection{Physics of Operation}
There are two types of hydrogen masers that serve as atomic frequency standards. The active hydrogen maser is the largest and most stable product on the market today, and the passive hydrogen maser is a smaller and lower-performance variation. The atomic transition involved is the ground-state hyperfine transition $\left|F=1, m_F=0\right> \rightarrow \left|F=1, m_F=1\right>$ of atomic hydrogen, with a transition frequency near 1420~MHz. Here, $F$ denotes the hyperfine energy level and $m_F$ denotes its sublevel.


In a hydrogen maser, a beam of atomic hydrogen is produced via RF excitation and directed into a cavity. Atoms are confined within a quartz bulb with a characteristic dimension less than the wavelength of microwave radiation (21~cm), satisfying the Dicke criterion \cite{dicke} and eliminating the first-order Doppler effect. An active hydrogen maser operates on the principle of self-sustained oscillation \cite{hellwig_review}; if the cavity losses are low enough and the intensity of the state-selected hydrogen beam is high enough, the collection of atoms and the microwave cavity interact in such a way as to sustain self-oscillation.

After microwave frequency synthesis, a quartz oscillator is phase-locked to this microwave frequency. A passive hydrogen maser operates on the same basic principle, but with sub-threshold gain of the cavity. In this case, a smaller storage bulb and/or lower-Q cavity are used, resulting in a smaller size device with comparatively poorer performance.

\subsubsection{Product Description and Performance}
Active hydrogen masers are the on the order of 300~L in volume and have the best performance in terms of short-term instability, phase noise, and flicker floor of any established products on the market. In the short term (on the order of 1-10~s), their $\sigma_y(\tau)$ degrades as $\tau^{-1}$, which is characteristic of white phase noise. This is due to the fact that in an active device the LO can be phase-locked, rather than frequency-locked, to the appropriate atomic transition. As can be seen in Fig.~\ref{ADEV_summary}, all other frequency standards shown are characterized by a $\sigma_y(\tau)$ that degrades as $\tau^{-1/2}$ in the short term, characteristic of white frequency noise. The intrinsic accuracy of a hydrogen maser is limited by the storage chamber wall properties, cavity detuning, and hydrogen density, which give rise to the long-term drift of these frequency standards. Commercial products exhibit a flicker floor of 10$^{-15}$ or less and a long-term drift of $2 \times 10^{-16}$/day. At present, hydrogen masers are available from three vendors: Microchip (USA), T4Science (Switzerland, also available through Orolia), and Vremya-CH (Russia).

An active hydrogen maser is the instrument of choice for applications requiring exquisite short-term stability such as radio astronomy and frequency metrology. One application of hydrogen masers is in maintaining precision phase stability among widely separated telescopes for Very Long Baseline Interferometry (VLBI), which were recently used in the Event Horizon Telescope to capture an image of a black hole~\cite{black_hole}.


\subsection{Cesium Beam Tube}
\subsubsection{Physics of Operation}
In Cesium Beam Tube (CBT) atomic frequency standards, the atomic transition involved is the ground-state hyperfine transition  $\left|F=3, m_F=0\right> \rightarrow \left|F=4, m_F=0\right>$ in~$^{133}$Cs, with a transition frequency of 9.192\hspace{0.8mm}631\hspace{0.8mm}770~GHz. These levels are referred to as ``clock levels," or ``field-independent transitions" and are used because they have no first-order dependence on external magnetic fields (only a small quadratic dependence).

Cesium is used in a beam tube frequency standard due to the fact that it has a high vapor pressure at reasonable oven temperatures, a small second-order Zeeman shift, a large hyperfine frequency, and atomic energy levels that are amenable to atomic state preparation, interrogation, and detection. Because of all these factors, it is also the basis of the SI definition of the second.

\begin{figure}[h]
	\centering
	\includegraphics[width=3.5in]{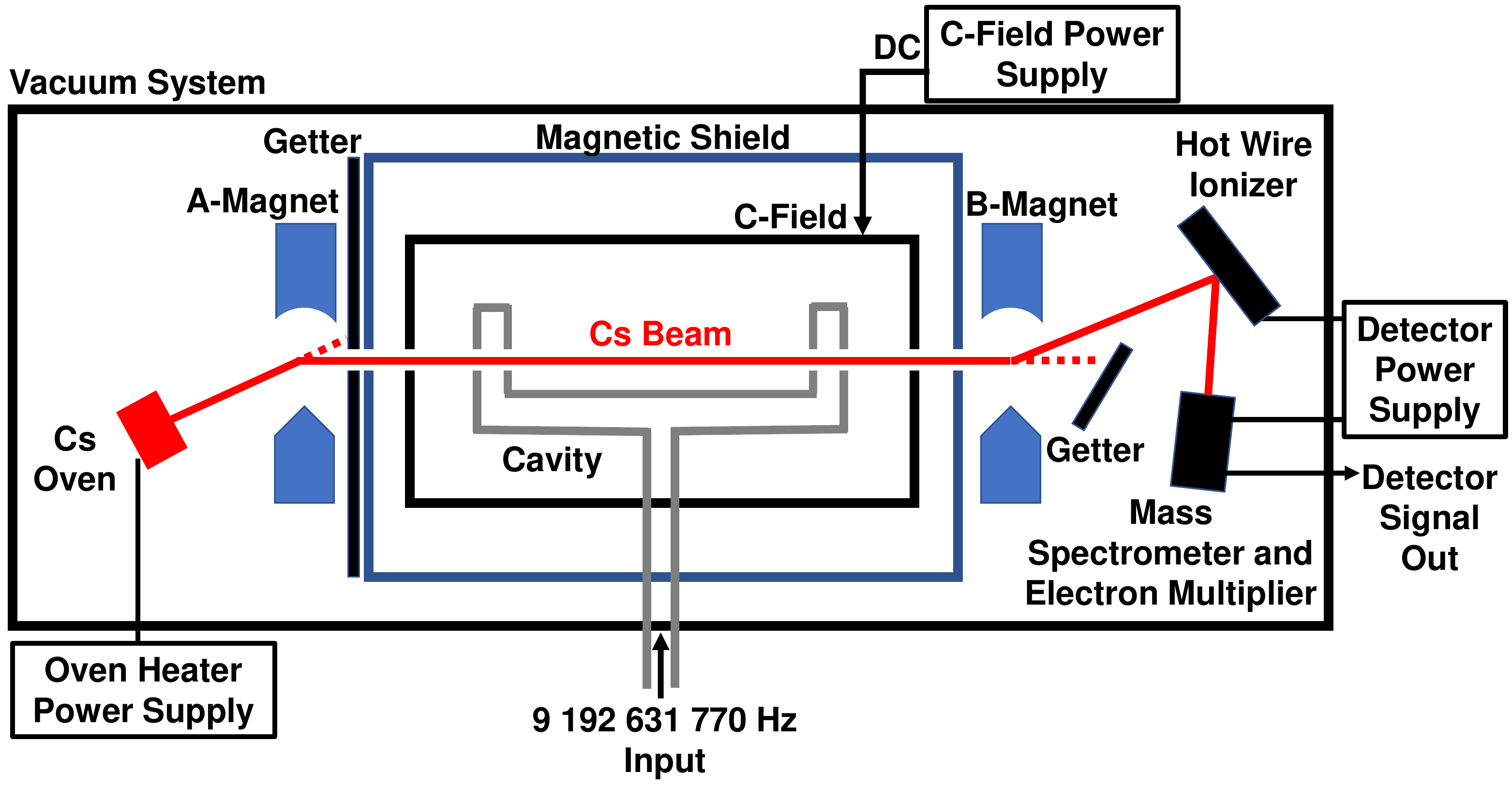} 
	\caption{Schematic of a Cesium Beam Tube atomic frequency standard (adapted from \cite{cutler50}). \label{CBT_schematic}}  
\end{figure}

A schematic  of a Cs beam tube frequency standard is shown in Fig.~\ref{CBT_schematic} \cite{Diddams_standards, cutler50}. Atoms effuse from a thermal oven towards the microwave cavity. The strong magnetic field generated by the A-Magnet deflects only one state into the U-shaped microwave cavity. Inside the microwave cavity, the microwaves cause a fraction of atoms to make a transition to the other clock state. Those atoms are then deflected by the B-Magnet onto a hot-wire detector and registered as signal.

\subsubsection{Product Description and Performance}
In commercial products \cite{cutler50}, atoms effuse from a thermal oven at a temperature on the order of 100~\textdegree C. Atoms pass through a Ramsey cavity with a length on the order of 10~cm, which results in a transit-time limited linewidth of $\sim$500~Hz and a line Q of $\sim2 \times 10^7$. The atomic linewidth is inversely proportional to the spacing between the two arms of the cavity. Based on Eq.~\ref{q_snr}, an SNR of $\sim$3000~$\sqrt{\text{Hz}}$ will result in an instability of $1 \times 10^{-11}$ at 1~s. The instability of commercial CBTs is limited by shot noise in the beam current \cite{cutler50}.

Accuracy limitations in commercial Cs beam tubes arise from an inability to exactly compensate for offsets in the measured clock frequency, including the C-field bias, the second-order Doppler shift, any residual phase difference between two arms of the cavity, and the servo system controlling the oscillator. These effects each contribute on the order of a few parts in $10^{13}$ to the device's specified accuracy \cite{audoin_review}.

\subsection{Gas Cell Frequency Standards}
Traditionally, the low-performance category of atomic frequency standards (typically based on Rb) have been categorized as ``Rb frequency standards.'' However, the introduction of cesium-based Coherent Population Trapping (CPT) frequency standards such as the Microchip CSAC into the commercial marketplace renders this name inaccurate. Because frequency standards of this category are all based on interrogation of a gas cell of atoms, we refer to this category as gas cell frequency standards. There are two main types of gas cell frequency standards: lamp-pumped rubidium vapor cell frequency standards, and  laser-pumped  CPT frequency standards. Both rely on an optical source (lamp or laser) for atomic state preparation, and both rely on the technique of microwave-optical double resonance (MODR). 

\subsubsection{Physics of Operation}
A gas cell frequency standard includes a vapor cell of alkali atoms (rubidium or cesium) that is simultaneously interrogated by both a light source and microwave radiation. In gas cell frequency standards, a light source is used for both atomic state preparation as well as state detection. The light source is responsible for preparing atoms in one of the two clock states via a process referred to as optical pumping. The ``double-resonance'' technique is named for the requirement that the optical and microwave fields are both resonant with the appropriate atomic transitions. When this condition is met, the atomic state population is transferred to a new energy state, which causes a change in how much light is absorbed in the gas cell and hence detected by a photodetector. The microwave radiation must be resonant with the transition frequency of the hyperfine clock levels (approximately 6.8~GHz in Rb or 9.2~GHz in Cs). The optical radiation is stabilized to an appropriate optical (ground-to-excited state) transition in the atoms and is used both to prepare the atomic state and to detect whether the microwaves are on resonance.

The core component of a gas cell frequency standard physics package is the gas cell (alternatively referred to as a vapor cell or alkali cell). This is a mm to cm-scale chamber made of a transparent glass such as quartz or borosilicate. Inside the vapor cell is a small amount of alkali metal. The gas cell is heated to a temperature on the order of $\sim$90~\textdegree C to increase the vapor pressure, resulting in measurable light absorption in the vapor cell. In order to reduce the frequency of depolarizing collisions with the chamber walls, an inert buffer gas is introduced into the chamber to slow down the collision rate. The buffer gas introduces a pressure shift on the clock transition, which in turn depends on the temperature of the buffer gas. Due to this environmental sensitivity as well as long-term drift from gas cell effects, these devices are not intrinsically accurate and need to be periodically calibrated.

\subsubsection{Physics of Lamp-Pumped Rb Frequency Standards}
A detailed description of the physics of operation of Rb frequency standards is provided in the literature \cite{lewis1991introduction, audoin_review}. In principle, the MODR technique can be achieved by optically pumping atoms into one of two clock states via a narrow-linewidth optical source. Traditionally RF discharge lamps, rather than lasers, are used for optical pumping and state detection because spectroscopic-grade lasers with the narrow linewidth and tunability required for integration into commercial products have only recently become available. The broadband lamp emission can be filtered into a narrow linewidth source via a scheme that was developed before the advent of the laser.

\begin{figure}[h]
	\centering
	\includegraphics[width=3.5in]{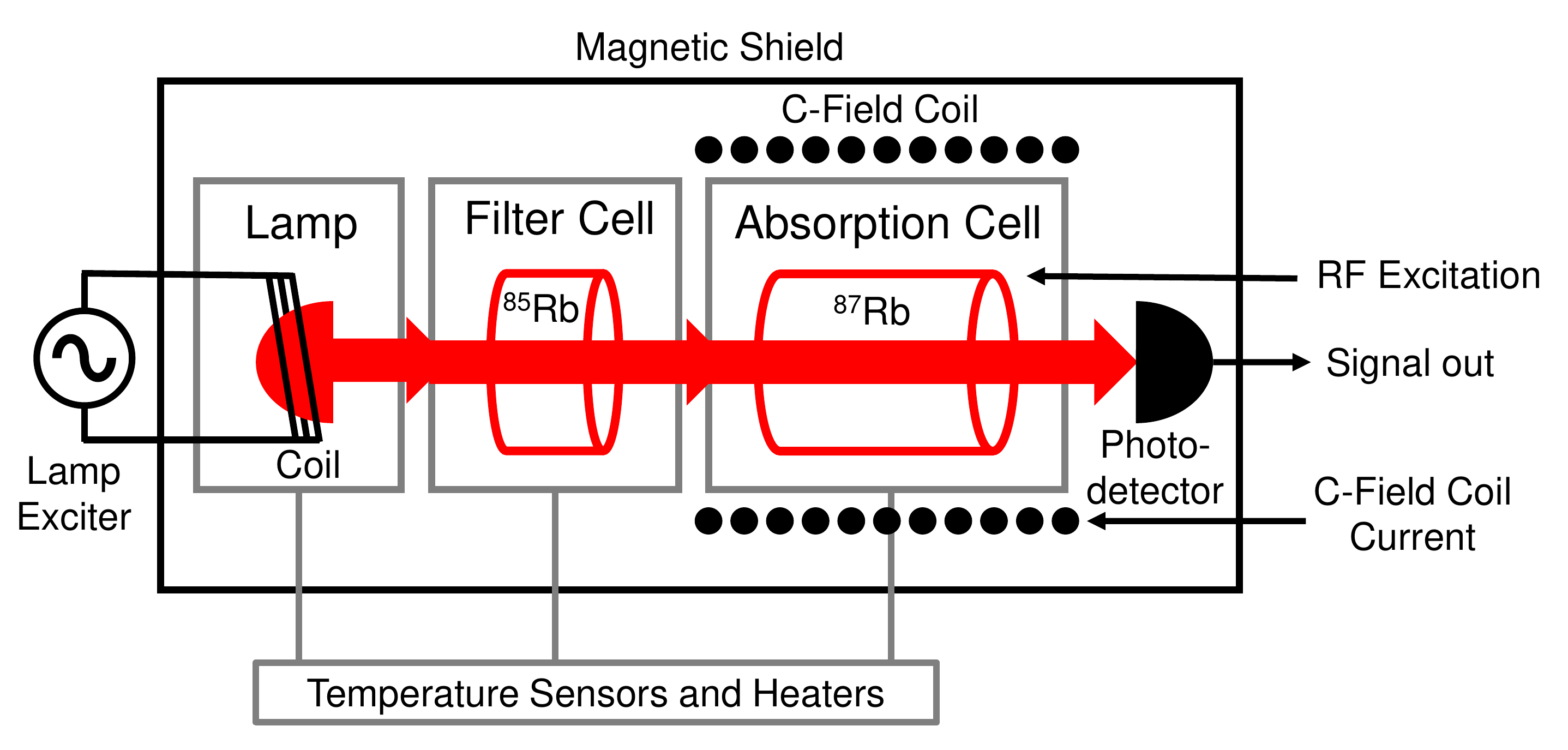} 
	\caption{Schematic of a lamp-pumped Rb frequency standard (adapted from \cite{riley_history}) where the output of a $^{87}$Rb lamp is first filtered by a $^{85}$Rb cell before entering the absorption cell. The optical output from the absorption cell is measured by a photodetector. \label{Rb_schematic}}  
\end{figure}

The atomic energy levels of $^{85}$Rb have a partial overlap with those of the $^{87}$Rb levels used in the frequency standard. This overlap can be exploited to generate laser-free optical pumping in an architecture shown in Fig.~\ref{Rb_schematic}. An RF discharge lamp made of $^{87}$Rb emits light at the resonances of $^{87}$Rb. This light is directed towards a filter cell made of $^{85}$Rb. Due to the fortuitous overlap of energy levels, one of these lines emitted from the $^{87}$Rb lamp is absorbed in the $^{85}$Rb filter cell. Therefore, light arriving at the $^{87}$Rb absorption cell, which is the core of the frequency standard, contains radiation at only one of the $^{87}$Rb (D1 transition) lines, which results in optical pumping into the $\left|F=2\right>$ state.

To accomplish MODR, microwave radiation is then applied to the absorption cell and transfers atoms from $\left|F=2\right>$ to $\left|F=1\right>$. This causes more atoms in the absorption cell to absorb the filtered light, which results in less light arriving at the photodetector. The electrical signal from the photodetector is then used as the input to the control loop for the frequency standard. In a simplified scheme, the filter cell is removed, resulting in distributed filtering within the absorption cell \cite{audoin_review}.

\subsubsection{Physics of Coherent Population Trapping}
\begin{figure}[h]
	\centering
	\includegraphics[width=3.5in]{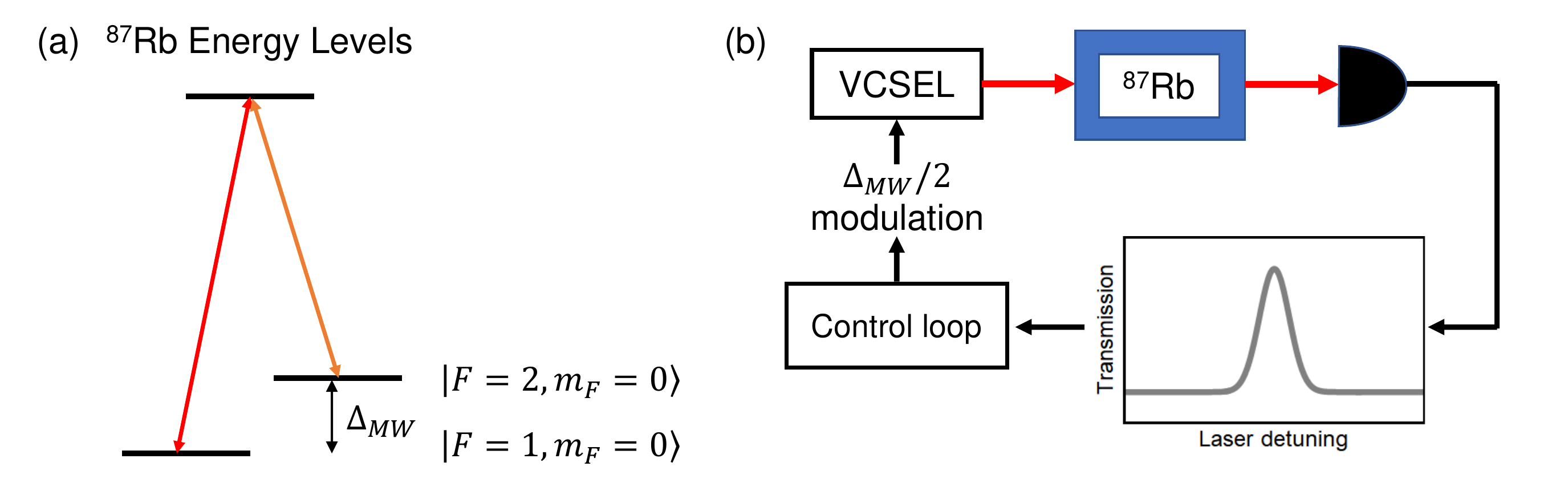} 
	\caption{Schematic of a CPT frequency standard (adapted from \cite{knappe_MEMS}). \label{CPT_schematic}}  
\end{figure}

The principles of Coherent Population Trapping \cite{cyr1993all} and its use in chip-scale atomic clocks (CSACs) have been covered in several reviews in the literature \cite{vanier_cpt, shah2010advances, knappe_MEMS}. In principle, CPT is similar to the underlying physical mechanism of lamp-based Rb frequency standards; an optical and a microwave source both need to be tuned to the appropriate transitions in alkali atoms, which generates a change in the amount of light incident on a photodetector. In practice, CPT is best understood as a three-level quantum mechanical interaction, as described in Ref.~\cite{knappe_MEMS} and illustrated in Fig.~\ref{CPT_schematic}.

When the two hyperfine ground states are simultaneously resonantly excited by two coherent optical frequencies, a type of destructive interference phenomenon occurs. Atoms are pumped into a so-called dark state, and there is an increase in optical transmission through the vapor cell. In practice, rather than two laser sources, a single laser source is used which is modulated at exactly half the hyperfine frequency. Atoms experience the laser modulation as an effective frequency detuning, which induces the CPT phenomenon. In this way, the microwaves are applied to the atoms indirectly via fast modulation of the laser current. This eliminates the need for a tuned microwave cavity and further simplifies the design of the physics package.

In practice, the technique of CPT is particularly amenable to miniaturization because the clock physics package can consist primarily of a single laser and a millimeter-scale microfabricated vapor cell. CSAC is based on an extreme example of miniaturization that includes a vertical cavity surface emitting laser (VCSEL) along with a Micro-Electro-Mechanical System (MEMS) microfabricated vapor cell \cite{lutwak2004chip}.

\subsubsection{Product Description and Performance: Lamp-Pumped Rb Frequency Standards}
Rb frequency standards are the workhorses of the telecommunications and tactical military markets, and there are more variations of this type of frequency standard on the market than any other frequency standard architecture. Rb frequency standard products vary and include exquisite devices currently used in the GPS constellation, rack-mounted devices in the telecom industry, and handheld units that consume a few Watts of power. 

\subsubsection{Product Description and Performance: CPT Frequency Standards}
The first CSAC physics package was demonstrated by NIST in 2004 \cite{knappe2004microfabricated}, and the first commercial CSAC was introduced by Microchip (then Symmetricom) in 2011. Now, there are a handful of products on the market today based on CPT. Some of these are advertised as CSACs with a power consumption of a few hundred mW. However, some CPT frequency standards are larger devices with a power consumption on the order of a few Watts.


\subsection{Comparison of Current Atomic Frequency Standards} \label{comparison}
A comparison of specifications from many current commercially available frequency standards is shown in Table~\ref{summary_table}. We have included long-established commercial products from companies such as Microchip, Accubeat, FEI, Spectratime, SRS, Excelitas, IQD, and Vremya. For the purposes of comparison, we have also included recent and emerging products from companies such as Teledyne, Chengdu Spaceon, Muquans, and Spectradynamics.

\begin{table*}[!h]
	\renewcommand{\arraystretch}{1.6}
	\caption[!t]{Summary of current frequency standard product key performance parameters. Tempco across T range is listed.}
	\label{summary_table}
		\fontsize{0.31cm}{0.32cm}\selectfont
		\begin{tabular}{>{\raggedright\arraybackslash}m{1.9cm}>{\raggedright\arraybackslash}p{1.4cm}>{\raggedleft\arraybackslash}p{1.5cm}>{\raggedleft\arraybackslash}p{0.9cm}>{\raggedleft\arraybackslash}p{1.3cm}>{\raggedleft\arraybackslash}p{1.3cm}>{\raggedleft\arraybackslash}p{1.2cm}>{\raggedleft\arraybackslash}p{1.3cm}>{\raggedleft\arraybackslash}p{1cm}>{\raggedleft\arraybackslash}p{0.8cm}>{\raggedleft\arraybackslash}p{0.7cm}}
			\textbf{Vendor} & {\textbf{Product}} & \centering{\textbf{ADEV}} \centering{($\tau=1$~s)} & \centering{\textbf{$\mathscr{L}$} (dBc/Hz, 10~Hz offset)} & \centering{\textbf{Drift} (monthly)} & \centering{\textbf{Retrace ($\pm$)}} & \centering{\textbf{T range}} \centering{$\left(^\circ\text{C}\right)$} & \centering{\textbf{Tempco}} & \centering{\textbf{Size}} \centering{(cm$^3$)} & \centering{\textbf{Weight} (kg)} & {\textbf{Power} (W)~} \\ 
			Microchip   & SA45.s CSAC & $3\times10^{-10}$     & -70      & $9\times10^{-10}$             & $5\times10^{-10}$    & -10 to 70            & $1\times10^{-9}$    & 17       & 0.035   & 0.12            \\
			Microchip   & SA35.m MAC  & $3\times10^{-11}$     & -87      & $1\times10^{-10}$             & $5\times10^{-11}$    & 0 to 75            & $7\times10^{-11}$ & 50        & 0.086 & 5                \\
			Microchip   & SA22.c      & $3\times10^{-11}$     & -90      & $4\times10^{-11}$             & $2\times10^{-11}$    & -10 to 75            & $1\times10^{-10}$    & 208      & 0.43  & 10              \\
			Microchip   & 5071A       & $5\times10^{-12}$     & -130     &                   &          & 0 to 55            &          & 29700     & 30   & 50               \\
			Microchip   & CsIII 4310B & $1.2\times10^{-11}$   & -130     &            &          & 0 to 50            &          & 16544      & 13.5 & 30               \\
			Microchip   & MHM         & $8\times10^{-14}$     & -138     & $6\times10^{-14}$           &                   &               &          & 371000      & 246 & 75               \\
			Accubeat    & NAC         & $2\times10^{-10}$     & -86      & $3\times10^{-10}$            &          & -20 to 65            & $2\times10^{-9}$    & 32        & 0.075  & 1.2             \\
			Accubeat    & AR133A      & $5\times10^{-12}$     & -116     & $1\times10^{-11}$            & $5\times10^{-11}$    & -20 to 65            & $1\times10^{-10}$    & 146       & 0.295  & 8.25            \\
			FEI         & FE-5669     & $6\times10^{-12}$  & -140     & $1\times10^{-11}$             & $2\times10^{-11}$ & -20 to 60            & $5\times10^{-11}$ & 669      & 1.69   & 20               \\
			FEI         & FEI RAFS    & $6\times10^{-13}$     & -138     &            & $5\times10^{-12}$    & -4 to 25            &          & 4902      & 7.5  & 39                \\
			Spectratime & LP Rb       & $1\times10^{-11}$     & -100     & $3\times10^{-11}$            & $5\times10^{-11}$    & -25 to 55            & $2\times10^{-10}$    & 216      & 0.29  & 10                \\
			Spectratime & iSpace RAFS & $3\times10^{-12}$     & -120     &              &          & -5 to 10            &          & 3224      & 3.4 & 35                 \\
			Spectratime & miniRAFS    & $1\times10^{-11}$     & -84      &           &          & -15 to 55            &          & 388      & 0.45  & 10                \\
			T4Science   & iMaser-3000 & $6\times10^{-14}$     & -136     & $6\times10^{-15}$             &                   &               &          & 436800     & 100       & 100        \\
			T4Science       & pHMaser     & $5\times10^{-13}$ & -130 &         &          &      &          & 49820   & 33 & 90      \\
			SRS             & PRS10       & $2\times10^{-11}$    & -130 & $5\times10^{-11}$       & $5\times10^{-11}$    & -20 to 65 & $2\times10^{-10}$    & 155   & 0.6 & 14.4    \\
			Excelitas       & RAFS        & $2\times10^{-12}$    & -105 & $3\times10^{-12}$     & $5\times10^{-12}$    & -20 to 45 &          & 1645   & 6.35 & 39    \\
			IQD             & IQRB-1      & $8\times10^{-11}$ & -95  & $5\times10^{-11}$         & $2\times10^{-11}$ & 0 to 50 & $5\times10^{-10}$ & 66     & 0.105 & 6     \\
			IQD             & IQRB-2      & $2\times10^{-12}$ & -138 & $4\times10^{-11}$        & $2\times10^{-11}$ &      &          & 230   & 0.22 & 6      \\
			Vremya          & VCH-1003M   & $6\times10^{-14}$ & -135 & $9\times10^{-15}$ &          &      &          & 305525 & 100 & 100    \\
			Chengdu Spaceon & XHTF1031 Rb & $5\times10^{-11}$ & -95  & $5\times10^{-11}$  &          & -30 to 65 & $2\times10^{-10}$ & 65     & 0.2 & 6      \\
			Chengdu Spaceon & XHTF1021 Rb & $3\times10^{-11}$ & -100 & $5\times10^{-11}$  & $2\times10^{-11}$ & -20 to 60 & $3\times10^{-10}$ & 189   & 0.27 & 7.8    \\
			Chengdu Spaceon & TA1000 OPC  & $1.2\times10^{-11}$ & -125 &          &          &     &          & 48266  & 40 & 100      \\
			Chengdu Spaceon & CPT         & $2\times10^{-10}$ & -90  &        & $5\times10^{-11}$ & -45 to 70 & $5\times10^{-10}$ & 24    & 0.045 & 1.6 \\
			Teledyne        & TCSAC       & $3\times10^{-10}$ & -85  & $3\times10^{-10}$  & $3\times10^{-10}$ & -10 to 60 & $1\times10^{-9}$ & 23      &    0.042 & 0.18   \\
			Muquans         & MuClock     & $3\times10^{-13}$ & -151 &           &          &      &          & 682000 & 135 & 200    \\
			Spectradynamics & cRb        & $5\times10^{-13}$ & -138 &        &          &      &          & 39806  & 30.5 & 75  
		\end{tabular}
\end{table*}

\begin{figure*}
	\centering
	\includegraphics[trim=0.2in 2.5in 0.2in 2.0in, clip, width=7in]{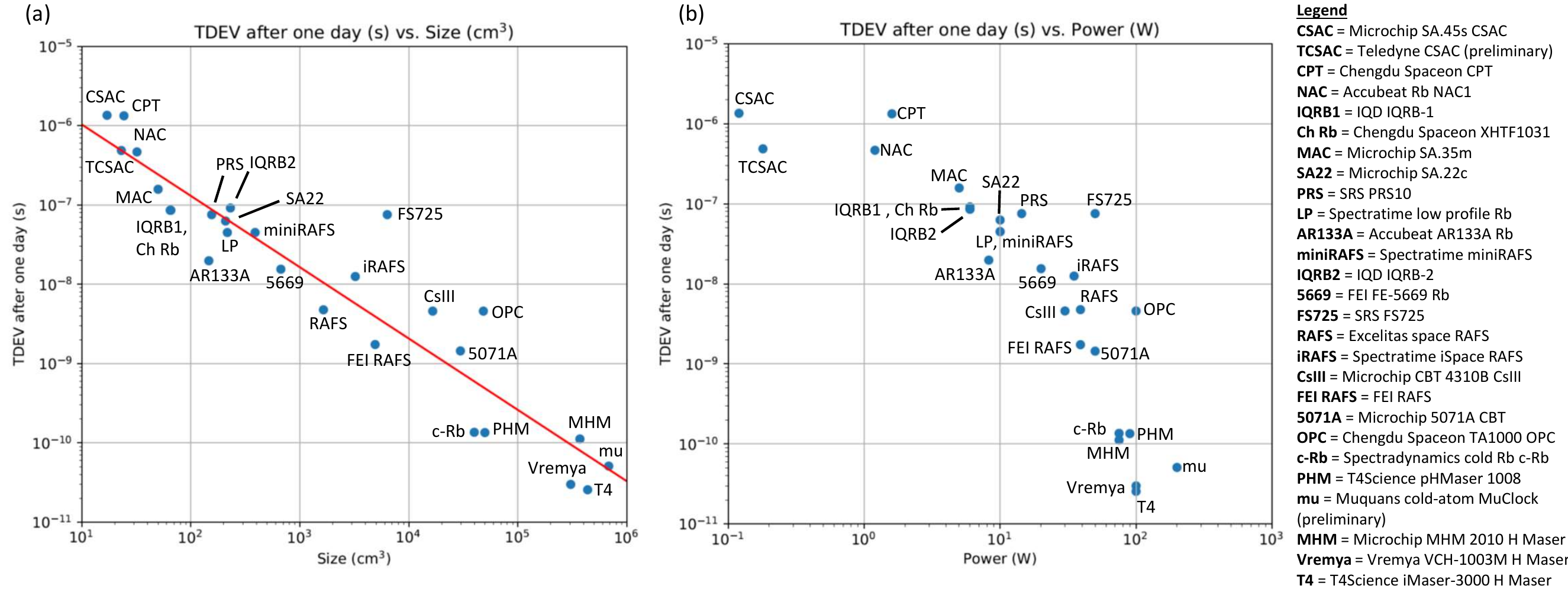} 
	\caption{TDEV at 1 day vs. (a) size and (b) power for the frequency standards listed in the legend. \label{TDEV_summary}}  
\end{figure*}

We generated figures showing the trends of ADEV ()$\sigma_y(\tau)$), $\mathscr{L}(f)$, drift, retrace,  tempco, and TDEV for all the frequency standards listed in Table~\ref{summary_table} as functions of size, weight, power, and the combined metric SWaP (a linear product of size, weight, and power), summarized in Ref.~\cite{cgsic}. We found that many of the observed trends are repetitive, and thus in the interest of space, here we show only TDEV after 1~day vs. size and power in Figs.~\ref{TDEV_summary}(a) and (b), respectively. TDEV after 1~day was chosen because it is a compromise between short-term and long-term behavior, and it incorporates both the frequency standard's ADEV and monthly drift specification. To calculate TDEV at 1~day, Eq.~\ref{time_error} was used, neglecting initial time error, frequency error, and environmental effects. If the product did not quote ADEV at 1 day, the contribution from ADEV was extrapolated as $\tau^{-1/2}$ from the largest $\tau$ value listed on the data sheet. The monthly drift specification was also extrapolated to 1 day. In practice, TDEV after 1~day for low-performance clocks is dominated by the drift spec.

The estimated TDEV(1 day) vs. size shown in Fig.~\ref{TDEV_summary}(a) fits well to a straight line on a log-log plot, indicative of a power-law relationship. This scaling relationship is also visible in plots of $\sigma_y(1~\text{s})$ vs. size, monthly drift vs. size, $\sigma_y(1~\text{s})$ vs. weight, and monthly drift vs. weight~\cite{cgsic}. The straight-line fit is shown in Fig.~\ref{TDEV_summary}, this is determined empirically to be $\text{TDEV(1 day)} = \mathcal{C}\times(\text{size in cc})^{\text{slope}}$; which yields $\mathcal{C}=8.17 \times 10^{-6}$ and slope of $-0.90$. This trend persists across five orders of magnitude in size and encompasses comparatively old technologies as well as modern CPT-based devices, including CSAC, low-performance Rb frequency standards, high-performance Rb frequency standards, CBT frequency standards, passive and active hydrogen masers, and emerging technologies such as cold atom frequency standards. This indicates empirically that an improvement in product performance can be expected from larger (or heavier) products according to a known power-law scaling.


The TDEV at 1 day vs. steady-state power consumption plot is shown in Fig.~\ref{TDEV_summary}(b). In contrast to the trend of TDEV vs. size, a saturation effect appears to be involved here, where gains in performance are no longer correlated with an increase in steady-state power consumption beyond 50-100~W. We note trends of $\sigma_y(1~\text{s})$ vs. power and monthly drift vs. power exhibit the same saturation effect~\cite{cgsic}.

Because of the redundancy involved in comparisons vs. size and weight, we conclude that analyzing performance parameters such as TDEV vs. the combined metric SWaP is not optimal. We conclude that comparisons of performance parameters vs. size and power separately are the most useful.

\section{Next-Generation Atomic Frequency Standards} \label{next_generation}
Despite ongoing advances in the performance and miniaturization of atomic frequency standards, portable microwave frequency standards appear to be approaching a practical limit due to the performance of quartz oscillators. It is difficult to find a volume source of quartz crystal oscillators with a short-term instability of better than $\sim1\times~10^{-13}$ at 1~s. By integrating a quartz LO and a microwave physics package with a clock loop tau on the order of $\sim$1~s, the resulting instability of the locked atomic frequency standard is limited to a slope on the order of $1 \times 10^{-13}/\sqrt{\tau}$. For performance beyond this limit, it is necessary to either radically improve quartz oscillators or employ an architecture based on a fundamentally different LO and physics package, \textit{e.g.}, an optical transition. In this section, we describe a range of next-generation microwave frequency standards followed by an overview of optical frequency standards under development. We analyze their current and projected performance relative to current frequency standard capabilities.

\subsection{Future practical microwave frequency standards}
There are a number of efforts underway to improve the performance of portable microwave frequency standards, which include both modifications to existing warm atom systems as well as the development of ion and cold atom-based frequency standards. Each of these platforms has exhibited short-term instabilities approaching $10^{-13}/\sqrt{\tau}$, as shown in Table~\ref{tableMW}. This level of instability is sufficient to enable applications such as deep-space autonomous navigation~\cite{7437483} and measurements of the gravitational redshift with an accuracy of 2 ppm~\cite{LAURENT2015540}. In this section, we will provide an overview of each of these platforms in terms of current performance and outlook for miniaturization.

\begin{figure*}
	\centering
	\includegraphics[width=6in]{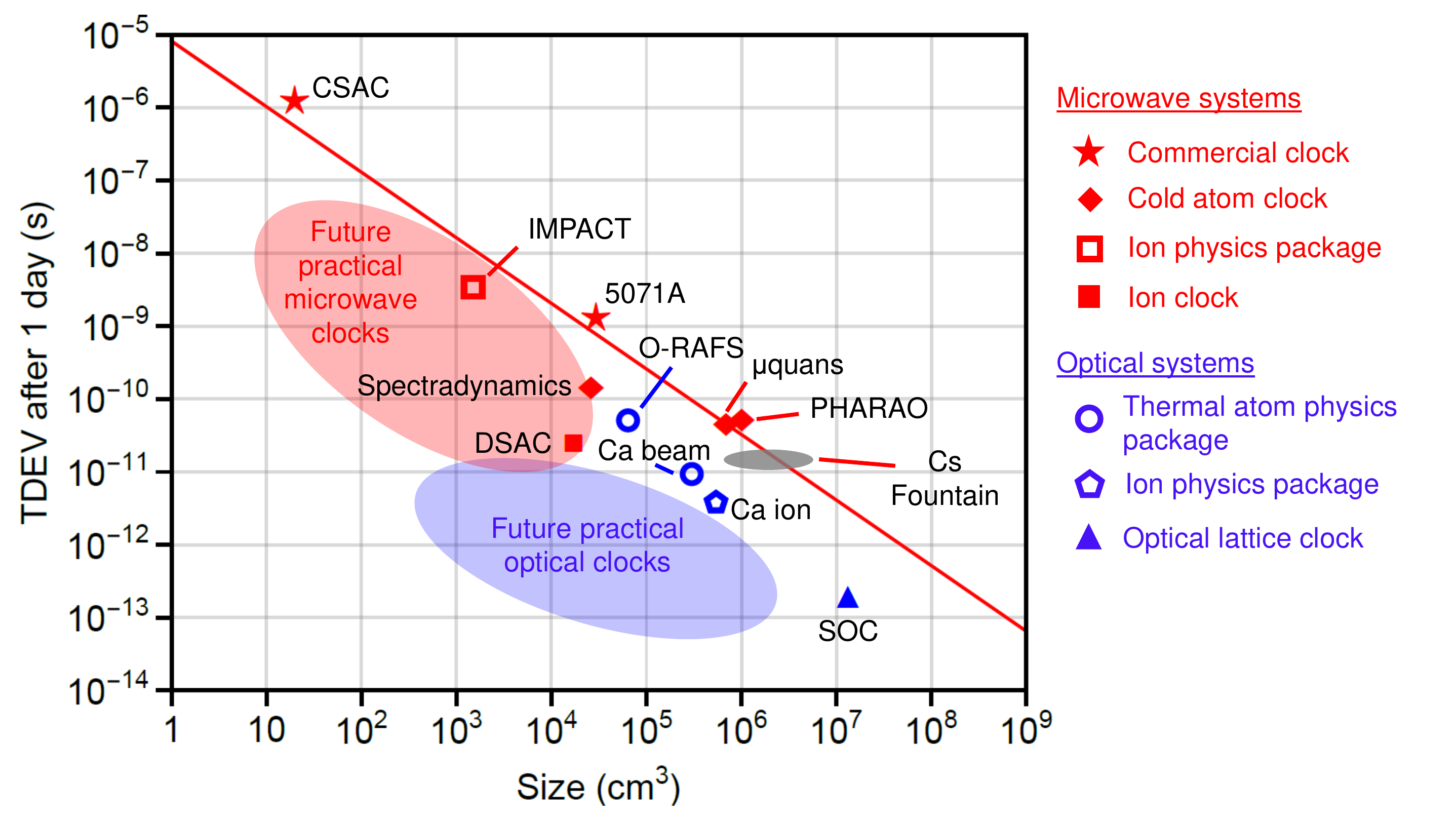}
	\caption{TDEV after 1 day vs. size for next-generation microwave and optical clocks. Solid filled symbols indicate clocks whose size includes the electronics package. Empty filled symbols indicate clocks whose size only includes the physics package. The red solid line is the fit TDEV($1\text{ day}$) = $8.17\times10^{-6}\text{(size in cc)}^{-0.9}$ used in Fig.~\ref{TDEV_summary}(a). The established commercial clocks (stars) are the Microchip CSAC and 5071A. The remainder of clocks shown are prototypes, technology demonstrations, or early-stage products. The cold atom microwave clocks are from SpectraDynamics~\cite{8597585, ascarrunz_PTTI}, Muquans~\cite{muquansclockdatasheet}, and projections for the PHARAO payload~\cite{LAURENT2015540}. The ion microwave clock is the JPL DSAC~\cite{7437483}. The ion microwave physics package is from IMPACT (Sandia)~\cite{7138951} (reported breadboard size $10\times15$~cm$^2$, assumed third dimension size of 10~cm). The thermal atom optical clock physics packages are AFRL O-RAFS~\cite{PhysRevApplied.9.014019, twophotonionpaper} (with size estimated from Ref.~\cite{8088927}) and the calcium thermal beam clock from Peking University~\cite{Shang:17}. The compact ion clock physics package is from the Chinese Academy of Sciences~\cite{wuhan1}. The optical lattice clock for the Space Optical Clock (SOC) program was developed by PTB and fits inside a trailer~\cite{PhysRevLett.118.073601}. The gray bubble describes current microwave fountains with parameters estimated from Refs.~\cite{1573940, fountainclocksPTBpress, Weyers_2018}. Red and blue bubbles are projections for future microwave and optical clocks, respectively, assuming further advances in stability and miniaturization.}
	\label{futureclockfigure}
\end{figure*}

Many high-performance microwave frequency standards have already undergone miniaturization efforts, and their ADEV at 1 second vs.~size is shown in Fig.~\ref{futureclockfigure}. Technology demonstrations that have reached the prototype stage are represented by closed symbols with the full size of the frequency standard included. For other advanced laboratory-based development efforts, only the approximate size of the physics package is included (represented by open symbols), with the assumption that future prototype frequency standards based on these technologies will undergo further miniaturization. The solid red curve in Fig.~\ref{futureclockfigure} shows the best fit of the TDEV vs.~size of current microwave frequency standards from Fig.~\ref{TDEV_summary}(a).

\subsubsection{Warm vapor microwave frequency standards}
Owing to the implementation of novel optical pumping techniques and increasingly stable narrow-linewidth semiconductor lasers~\cite{Hafiz_2016, PhysRevApplied.7.014018}, warm atomic vapor-based microwave frequency standards have achieved substantial performance improvements in recent years, as shown in Table~\ref{tableMW}. 

Recent demonstrations have achieved instabilities that rival some portable hydrogen masers~\cite{PhysRevApplied.7.014018, Gharavipour_2016}. Although these systems have not yet undergone explicit miniaturization, many engineering hurdles have already been overcome with the development of CSACs, and thus this platform holds promise for future frequency standard applications that require ultra-low SWaP.

\subsubsection{Cold atom microwave frequency standards}
\begin{figure}
	\begin{center}
		\includegraphics[width=3.5in]{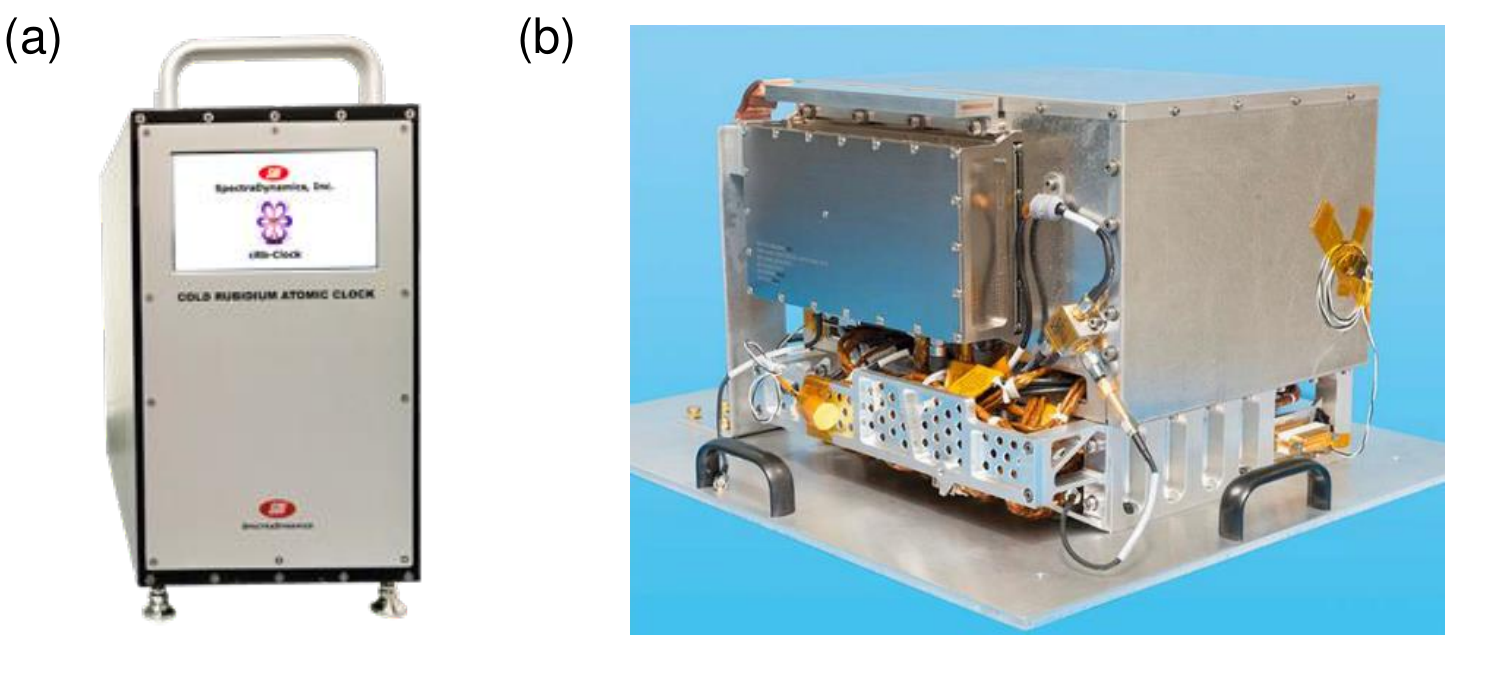}
		\caption{(a) The Spectradynamics cold atom clock (height = 48~cm) reproduced with permission from Ref.~\cite{spectrawebsite}. (b)~The DSAC system (volume = 17~L) from Ref.~\cite{7437483}.}
		\label{microwaveclockimages}
	\end{center}
\end{figure}

Efforts to reduce the coupling of atoms to the external environment have motivated the development of neutral cold-atom systems, which exhibit much smaller Doppler and collisional shifts than warm atoms and can have a smaller SWaP than atomic fountains. In these systems, atoms are first laser-cooled and trapped in a magneto-optical trap (MOT). This both cools the atoms, slowing them down to speeds on the order of 10~cm/s or less, and confines them to a well-defined region using magnetic field gradients. In recent years, there have been significant efforts towards miniaturization of these systems \cite{rushton2014contributed}. This type of architecture requires a periodic timing sequence with periods of trapping, laser cooling, clock interrogation with the optical and magnetic trapping field turned off, and state readout. Because atoms are untrapped (and hence free falling due to gravity), the maximum practical duration of the clock interrogation phase is limited to 10-100~ms, which sets limits on frequency standard stability. Compact versions of cold atom frequency standards have achieved short-term instabilities as low as $\lesssim3\times10^{-13}/\sqrt{\tau}$, as shown in Table~\ref{tableMW}, and it is expected that the PHARAO clocks on the ACES project of the European Space Agency (ESA) will reach $\lesssim1\times10^{-13}/\sqrt{\tau}$~\cite{LAURENT2015540}. Compact cold atom frequency standards are already operating in orbit~\cite{Liu2018} and have flown on an aircraft~\cite{PhysRevApplied.10.064007}, and they are commercially available from Muquans~\cite{muquansclockdatasheet} and SpectraDynamics~\cite{ascarrunz_PTTI}, with the latter system depicted in Fig.~\ref{microwaveclockimages}(a).

A plot of TDEV(1 day) vs. size for these cold atom clocks is shown in Fig.~\ref{futureclockfigure}. For future practical microwave frequency standards, we include advanced prototype demonstrations with some level of transportability, an overall size of the frequency standard or physics package listed, and a measured ADEV out to at least $10^4$~seconds. For future practical optical frequency standards, we include those demonstrations matching the conditions above but with a measured ADEV out to at least $10^3$~seconds, and we calculate TDEV in the same way.

From Fig.~\ref{futureclockfigure} it is clear that the TDEV(1 day) vs.~size of these emerging products and next-generation prototypes follows a similar relationship as current frequency standards, which suggests that they are not achieving stability improvements for a given size compared to today's technology. 

\subsubsection{Trapped ion microwave frequency standards}
Ion-based systems are also making substantial progress in SWaP reduction~\cite{8597512}. Ions that are used in frequency standards are positively charged, \textit{i.e.}, they are atoms that are missing an electron. Ions can be trapped by applying certain configurations of electric and/or magnetic fields that confine charged particles. Ion traps can consist of a single or few ions or a cloud of up to approximately 10 million ions. Single ions trapped in deep potential wells operate under exquisite conditions that are well-isolated from many external perturbations. As a result, even though there can be far fewer ions used in ion-based frequency standards than there are atoms in vapor-based frequency standards, they can still provide a large SNR. In addition, ion traps are highly stable with lifetimes of up to months, thus enabling long interaction times during clock interrogation, whereas cold atoms must be continually re-cooled and trapped (typically every 10-100~ms). The ion platform is under development using a number of atomic species, as shown in Table~\ref{tableMW}, with one of the most advanced being the mercury Deep-Space Atomic Clock (DSAC) depicted in Fig.~\ref{microwaveclockimages}(b).

The DSAC developed by the Jet Propulsion Laboratory (JPL) and NASA has reached an \emph{SNR}~$\times$~\emph{Q}-limited short-term instability of $1.5\times10^{-13}/\sqrt{\tau}$~\cite{7437483}. The ion interrogation cycle begins by applying light from a mercury lamp to optically pump the trapped ions into a magnetically insensitive ground state. A $40.5$ GHz microwave field is then applied to the ions, which gives rise to enhanced rates of fluorescence when its frequency is tuned to resonance with the hyperfine splitting. The fluorescence is collected with a photomultiplier tube~\cite{7437483}, the output of which is used to generate an error signal that feeds back to the microwave source. The DSAC system is well-suited for long-term operation in space because it uses a lamp instead of a laser and does not require cryogenics or a microwave cavity.

The DSAC performs better than the current product trendline for its size, \textit{i.e.}, compared to the red solid curve in Fig.~\ref{futureclockfigure}. We expect that this system is near the limit of next-generation microwave frequency standard performance vs.~size, which we represent in Fig.~\ref{futureclockfigure} by the red bubble. We estimate the parameters of this bubble by noting that, while we expect some further advances in performance and size relative to current frequency standards (as demonstrated in the DSAC, for example), we do not expect future microwave frequency standards to reach sizes that are substantially smaller than current CSACs nor for practical frequency standards to reach short-term instabilities much better than $10^{-13}$ at 1 second (limited by the instability of high-performance quartz oscillators). Some high-performance microwave ion systems have been shown to reach sub-$10^{-13}/\sqrt{\tau}$ instabilities, but these systems are generally more complex and require laser cooling of the ions or high-SWaP local oscillators (\textit{e.g.}, superconducting cavity masers~\cite{1439784}). To surpass the current trends in performance vs.~size of microwave frequency standards, it is necessary to utilize a completely different frequency standard platform with an intrinsically higher $Q$\textemdash one referenced to atomic transitions separated by optical, rather than microwave, frequencies.

\begin{table*}[!t]\centering
		\caption{Examples of microwave atomic frequency standards under development.}
		\renewcommand{\arraystretch}{1.4}
		\fontsize{0.31cm}{0.32cm}\selectfont
		\begin{tabular}{| >{\centering\arraybackslash}m{1.8cm} | >{\centering\arraybackslash}m{3cm} | >{\centering\arraybackslash}m{0.7cm} | >{\centering\arraybackslash}m{2.4cm} | >{\centering\arraybackslash}m{2.1cm} | >{\centering\arraybackslash}m{1.8cm} | >{\centering\arraybackslash}m{2.3cm} |}
			\hline
			\textbf{Platform} Advantages & \textbf{System information} & \textbf{Atom} & \textbf{Short-term instability} & \textbf{Long-term performance} & \textbf{Uncertainty} & \textbf{Performer(s)}\\ \hline
			
			&  CPT (Mclocks project result)  & Rb & $3.2\times10^{-13}/\sqrt{\tau}$ & Flicker floor = $3\times10^{-14}$ at 300~s, Drift = $1\times10^{-10}$/month & Not reported & LNE-SYRTE, INRIM, UFC~\cite{Hafiz_2016}, ~\cite{PhysRevApplied.7.014018} \\ \cline{2-7}
			
			\textbf{Warm vapor}\newline Projected low SWaP, High atomic densities & CPT, Physics package contained on 2.54~mm$\times$30~mm board & Rb & $7\times10^{-11}/\sqrt{\tau}$, 1 to 100~s & Flicker floor = $8\times10^{-12}$~(100-1000~s) & Not reported & Universit\'{e} de Neuch\^{a}tel, SpectraTime (SpT)~\cite{6533686} \\ \cline{2-7}
			
			&  Double-resonance pumping scheme  & Rb & $1.4\times10^{-13}/\sqrt{\tau}$, 1 to 100~s & Not reported & Not reported & Universit\'{e} de Neuch\^{a}tel~\cite{Gharavipour_2016} \\ \hline
			
			&  CAMPS program result  & Rb & $1.5\times10^{-10}/\sqrt{\tau}$, 1 to $10^{5}$~s & Not reported & Not reported & Honeywell~\cite{7546775} \\ \cline{2-7}
			
			&  Commercial cRb, 22$\times$37$\times$32 cm$^3$, 28 kg  & Rb & $8\times10^{-13}/\sqrt{\tau}$, 1 to $10^4$~s & Flicker floor $\approx9\times10^{-16}$ at $10^6$~s, Drift $<1\times10^{-17}$/day & Not reported & SpectraDynamics, NIST~\cite{8597585},~\cite{ascarrunz_PTTI} \\ \cline{2-7}
			
			\textbf{Cold vapor}\newline    Reduced Doppler and collisional effects & Commercial MuClock, 155$\times$55$\times$80~cm$^3$, 135~kg  & Rb & $3\times10^{-13}/\sqrt{\tau}$, 1 to $10^4$~s & Flicker floor $\approx2\times10^{-15}$ at 10 days & few parts in $10^{-15}$ & Muquans~\cite{muquansclockdatasheet} \\ \cline{2-7}
			
			&  Cold atom clock experiment in space (CACES) demonstrates tests in orbit & Rb & $2\times10^{-12}/\sqrt{\tau}$ measured on ground, 1 to $10^5$~s ($3\times10^{-13}/\sqrt{\tau}$ predicted in orbit) & Not reported & Not reported & Chinese Academy of Sciences~\cite{Liu2018} \\ \cline{2-7}
			
			&  HORACE clock for Galileo GNSS  & Cs & $2.2\times10^{-13}/\sqrt{\tau}$, 1 to $10^4$~s & Not reported & Not reported & LNE-SYRTE~\cite{ESNAULT2011854} \\ \cline{2-7}
			
			&  Projection for the ACES (PHARAO) payload  & Cs & $3.4\times10^{-13}/\sqrt{\tau}$ out to $3\times10^4$~s & Not reported & $1.4\times10^{-15}$ &  LNE-SYRTE, CNES, ESA, ENS-PSL Research University~\cite{LAURENT2015540}  \\
			\hline
			
			\textbf{Fountain}\newline  Well-established high performance &  Physics package designed for commercial development of fountains & Cs & $1\times10^{-13}/\sqrt{\tau}$, 1 to $10^4$~s & Not reported & Not reported & National Physical Laboratory, Astrogeodynamical Observatory, National Research Council of Canada, Penn State~\cite{commercialfountain} \\ \hline
			
			&  DSAC (Deep-space atomic clock), 17,000~cm$^3$, 16 kg, 47 W  & Hg & $1.5\times10^{-13}/\sqrt{\tau}$, 1 to $10^5$~s & Drift $<6\times10^{-16}$/day & Not reported & JPL, NASA~\cite{7437483} \\ \cline{2-7}
			
			\textbf{Trapped ion}\newline  Well-isolated from environment & Sympathetically cooled Cd ions and laser-cooled Mg ions in Paul trap & Cd & $6.1\times10^{-13}/\sqrt{\tau}$, 4 to $4\times10^3$~s & Not reported & $6\times10^{-14}$ & Tsinghua Univ.~\cite{Miao:15} \\ \cline{2-7}
			
			&  Ultra-small vacuum packages under development for the IMPACT program  & Yb & $2\times10^{-11}/\sqrt{\tau}$, 10 to $10^4$~s & Not reported & Not reported & Sandia, JPL~\cite{doi:10.1063/1.4948739} \\ \cline{2-7}

			&  Micro-Mercury Lamp-Pumped clock under development for the ACES program (30 cm$^2$ vacuum package)  & Yb & $1.7\times10^{-12}/\sqrt{\tau}$, 10 to 1000~s & Flicker floor = $7\times10^{-14}$ at 1000~s & Not reported & JPL~\cite{hoang2019performance} \\ \cline{2-7}
			
			&  Laser-cooled ion clock contained in volume of 51$\times$49$\times$28 cm$^3$\vspace{1mm} & Yb & $3.6\times10^{-12}/\sqrt{\tau}$, 30-1500~s & Not reported & Not reported & National Physical Laboratory, Univ. of Oxford~\cite{mulholland2019laser} \\ \hline
		\end{tabular}
		\label{tableMW}
\end{table*}

\subsection{Optical frequency standards}
Future compact frequency standards based on optical transitions are expected to reach short-term fractional frequency instabilities of approximately $1 \times 10^{-14}/\sqrt{\tau}$ to \mbox{$1 \times 10^{-15}/\sqrt{\tau}$~\cite{ludlow_optical}}. These frequency standards will enable longer-term autonomous navigation and improved phase precision for applications including distributed coherent radar, beamforming, and geolocation.


Optical frequency standards offer fundamentally improved performance over microwave frequency standards. This improvement can be understood from Eq.~\ref{q_snr}, where it is clear that employing a higher reference frequency (higher $Q$) improves the stability of the frequency standard. Optical frequency standards use an optical field ($\nu\sim$ hundreds of THz) as a frequency reference, which offers an immediate projected improvement in stability over microwave frequency standards (typically $\nu\sim$ 1-10 GHz) by a factor of $\sim$10,000. 

In practice, as is the case in microwave frequency standards, optical frequency standards can suffer from a number of systematic shifts that must be shielded against or corrected: Doppler, Zeeman, Stark, collisional, and gravitational effects can shift the energy levels of the atoms. In many cases, the magnitude of these shifts is inversely proportional to the clock frequency, which makes optical frequency standards intrinsically less susceptible to environmental factors than their microwave counterparts. 

The overall schematic of a generic optical frequency standard is shown in Fig.~\ref{clock_control_loop}(b). The feedback loop is in many ways analogous to that of a microwave frequency standard, but in this case, the LO is a laser (which often requires its own cavity for stabilization). When the frequency of the optical field is tuned close to a ground-to-excited state transition, the atoms/ions will absorb the light with a high probability. The absorption is maximized when the optical frequency is tuned precisely to the atomic resonance, and thus the output optical power recorded by a detector can be used to produce an error signal that is fed back to the laser system. A fraction of the stabilized laser output is also coupled into a frequency comb, which in this case acts as a frequency downconverter from the optical to the microwave domain and serves to produce the RF/microwave output of the frequency standard.

Optical frequency standards are expected to form the basis for a re-definition of the SI second~\cite{McGrew:19} because they have surpassed state-of-the-art microwave fountains in both uncertainty and instability. We note that there have been a number of recent review articles focused on optical atomic frequency standards~\cite{ludlow_optical, Riehle2017, Poli2013opticalrev}, which provide extensive overviews of optical frequency standard technologies and recent progress. Here we will provide brief descriptions of the range of optical frequency standards currently under development with a focus on progress towards building practical, portable optical frequency standards and their projected impact.

\subsubsection{Thermal atom optical frequency standards}
Thermal atoms remain a promising platform for a wide range of future practical optical frequency standards because they enable access to a larger number of atoms with a relatively simple physics package. Systems based on vapor cells and atomic beams are under development, some examples are outlined in Table~\ref{tableMW}. Warm vapor-based clocks must employ schemes to minimize Doppler effects, such as Doppler-free spectroscopy~\cite{keplerieee17} or, more recently, a two-photon transition in rubidium~\cite{PhysRevApplied.9.014019}. The demonstrated short-term instability of these systems is already better than a number of compact microwave frequency standards, as shown in Fig.~\ref{futureclockfigure}.

\begin{figure}
	\begin{center}
		\includegraphics[scale=0.6]{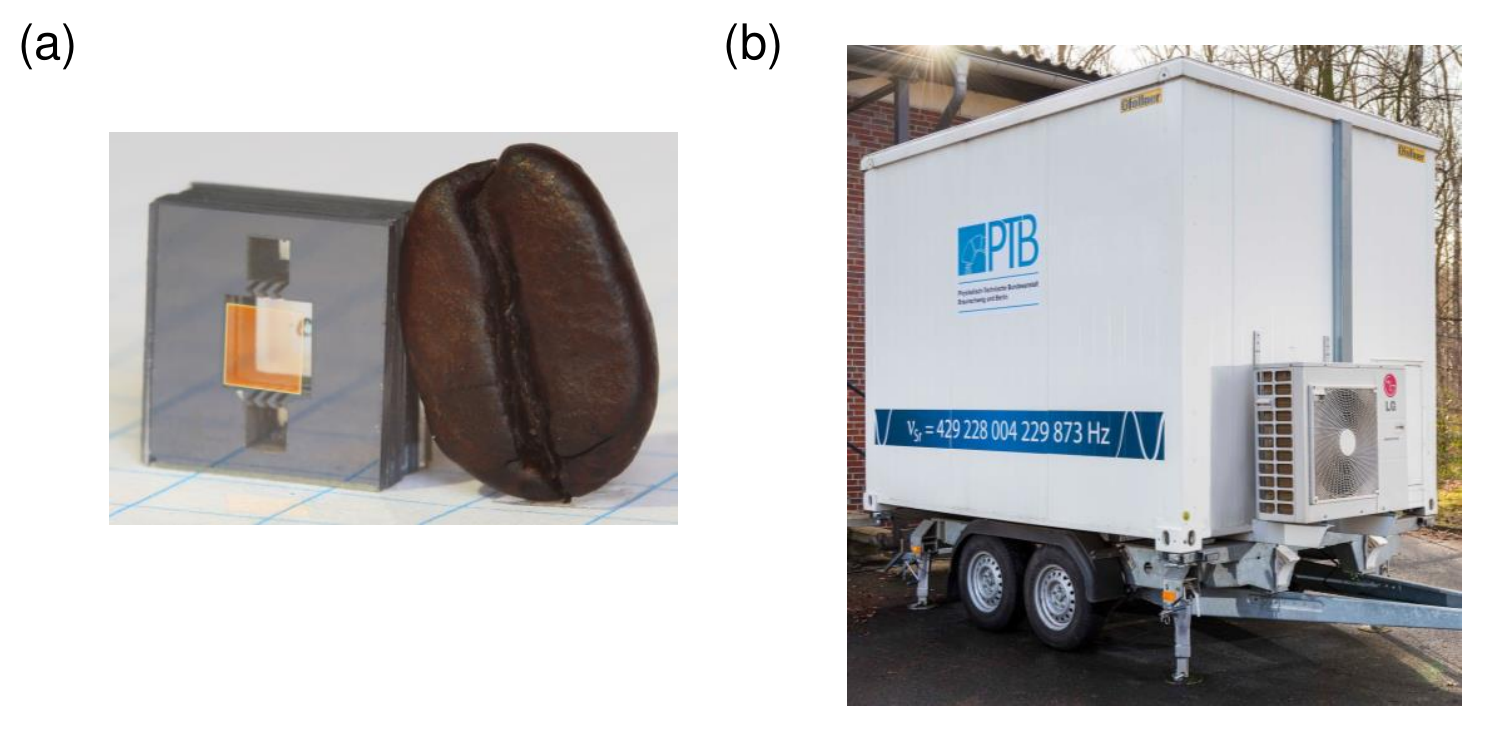}
		\caption{(a) The vapor cell on a chip used in the two-photon thermal atom optical frequency standard~\cite{Newman:19} (image from press release~\cite{nistopticalbeanclock}) shown next to a coffee bean. (b)~The PTB optical lattice frequency standard on an air-conditioned car trailer~\cite{PhysRevLett.118.073601}.
		}
		\label{opticalclockimages}
	\end{center}
\end{figure}

We expect thermal atom optical frequency standards, along with ion-based optical frequency standards, to eventually populate the smaller-size region of the ``Future practical optical frequency standards'' blue bubble in Fig.~\ref{futureclockfigure}, where technologies including ultra-small vapor cells, like the one shown in Fig.~\ref{opticalclockimages}(a), and chip-integrated frequency combs~\cite{Spencer2018} will enable substantial further miniaturization. 


Optical frequency standards based on thermal atomic beams can eliminate some collisional effects present in vapor cell systems and take advantage of some of the techniques used for cesium beam microwave frequency standards. Laboratory-scale optical atomic beam frequency standards have also reached fractional frequency instabilities of less than $10^{-14}$ at 1~s~\cite{6243750}, and miniaturization efforts are underway~\cite{Shang:17}. Even with first-order Doppler-free spectroscopy techniques, thermal beams still suffer from second-order Doppler effects~\cite{Poli2013opticalrev}, and thus a number of systems have begun to employ laser cooling and magneto-optical trapping.

Frequency standards based on freely expanding laser-cooled atoms can achieve longer coherence times and a negligible second-order Doppler shift compared to warm atom systems. However, technical challenges of optimizing the cooling schemes (\textit{e.g.}, mode mismatch between the cooling fields and the atoms) are expected to limit the performance of these frequency standards to fractional uncertainties of about $10^{-15}$~\cite{Poli2013opticalrev}. State-of-the-art setups have therefore moved to lattice-based schemes~\cite{McGrew2018}, and some future practical optical frequency standards, including the Space Optical Clocks (SOC) project~\cite{Origlia:17} of ESA, have also been commissioned owing to their promise of enhanced performance.

\subsubsection{Optical lattice frequency standards}
Counterpropagating optical fields applied to a vapor of cold neutral atoms form an ``optical lattice,'' which imparts a dipole force that attracts atoms to the minima of the dipole potential wells. For sufficiently cold atoms and deep potential wells, the atoms become spatially confined at the minima of the potential wells, each of which can be much smaller than the wavelength of light. Under conditions of very tight confinement, the light-atom system can enter the so-called ``Lamb-Dicke regime,'' in which Doppler effects and atomic recoil due to photon scattering are suppressed. In this regime, optical lattice-based frequency standards are highly insensitive to residual atomic motion and can surpass the performance of frequency standards which use freely expanding atoms.

Atoms that are exposed to strong optical fields experience internal energy shifts (Stark shifts). By employing a carefully chosen wavelength~\cite{PhysRevLett.91.173005}, referred to as the ``magic wavelength,'' some of those shifts can be canceled, leaving only residual effects such as those due to wavefront distortions. Under these conditions, laboratory optical lattice frequency standards have reached record levels of instability (ADEV of $4.7\times10^{-17}$ at 1~s) and fractional uncertainty ($5.8\times10^{-19}$ after 1 hour of averaging)~\cite{Oelker2019}. The SOC program is supporting the development of transportable optical lattice frequency standards (Sr and Yb) with goals of a fractional instability below $1\times10^{-15}/\sqrt{\tau}$ and a fractional uncertainty below $5\times10^{-17}$ for the next generation of frequency standards beyond PHARAO~\cite{Origlia:17}. 

To our knowledge, the Sr lattice frequency standard from PTB is currently the only full lattice frequency standard that has been constructed on a transportable platform~\cite{PhysRevLett.118.073601}, \textit{i.e.}, an air-conditioned trailer shown in Fig.~\ref{opticalclockimages}(b). Its fractional frequency instability is two orders of magnitude better than the cold-atom-based transportable microwave frequency standards shown in Fig.~\ref{futureclockfigure}. We expect that further miniaturization and integration of system components will shift this platform into the high-performance area of the blue bubble defining ``future practical optical frequency standards'' in Fig.~\ref{futureclockfigure}.


\subsubsection{Trapped ion optical frequency standards}
Despite having a lower number of atoms, trapped ion optical frequency standards represented the highest performance optical frequency standards for a number of years after the development of optical frequency combs, with optical lattice platforms only surpassing them in 2014~\cite{Bloom2014}. As is the case in microwave ion frequency standards, their excellent instability is due in large part to their pristine environment. In addition, laser-cooled ions can also operate in the Lamb-Dicke regime when using sufficiently deep trapping potentials. Even though optical lattice systems have surpassed that of ions in terms of performance, ion-based systems have much simpler physics packages than optical lattice systems and thus are expected to attain a much smaller SWaP, which makes them highly amenable to a number of strategic applications in navigation and communications. In addition, the long-lived ion traps can be exploited to obtain a higher \emph{Q} and SNR.

State-of-the-art ion optical frequency standards have achieved short-term fractional frequency instabilities nearing $10^{-15}/\sqrt{\tau}$ and fractional uncertainties of $9.4\times10^{-19}$~\cite{brewer}, as shown in Table~\ref{highperformanceclocks}. Efforts to miniaturize ion optical frequency standards are also underway~\cite{Hannig2019}, with one system reaching a volume of $0.54$~m$^3$ excluding electronics~\cite{wuhan1}. 

\begin{table*}[!t]\centering
		\fontsize{0.31cm}{0.32cm}\selectfont
		\caption[h]{Examples of optical atomic clocks under development.}
		\renewcommand{\arraystretch}{1.4}
		\begin{tabular}{| >{\centering\arraybackslash}m{2.6cm} | >{\centering\arraybackslash}m{4.0cm} | >{\centering\arraybackslash}m{0.8cm} | >{\centering\arraybackslash}m{2.4cm} | >{\centering\arraybackslash}m{2.0cm} | >{\centering\arraybackslash}m{2.7cm} |}
			\hline
			\textbf{Platform} Advantages  & \textbf{System information} & \textbf{Atom} & \textbf{Short-term instability} & \textbf{Uncertainty} & \textbf{Performer(s)}\\ \hline
			
			& O-RAFS - Two-photon transition scheme in vapor cell & Rb & $~~~3\times10^{-13}/\sqrt{\tau}$,\vspace{2mm} 1 to $10^4$~s & Not reported &  AFRL, NIST~\cite{PhysRevApplied.9.014019}, \cite{8088927}, \cite{twophotonionpaper} \\ \cline{2-6}
			
			\textbf{Warm atoms}\newline Low SWaP+C, High densities &  Two-photon transition scheme with vapor cell on a chip and microresonator comb  & Rb & $4.4\times10^{-12}/\sqrt{\tau}$,\vspace{2mm} 0.1 to $10^3$~s & Not reported & NIST, UC Boulder, Cal Tech, Draper, Stanford~\cite{Newman:19} \\ \cline{2-6}
			
			& Atomic beam, miniaturized physics package (0.3 m$^3$) & Ca & $5.5\times10^{-14}/\sqrt{\tau}$,\vspace{2mm} 0.1 to $10^3$~s & Not reported & Peking Univ., Beijing Vacuum Electronics Research Inst.~\cite{Shang:17} \\ \hline
			
			\textbf{Optical lattice}\newline Higher densities, insensitive to atomic motion &  Clock installed on air-conditioned trailer  & Sr & $1.3\times10^{-15}/\sqrt{\tau}$,\vspace{2mm} 3 to $2\times10^{3}$~s & $7.4\times10^{-17}$ & PTB~\cite{PhysRevLett.118.073601} \\ \hline
			
			\textbf{Trapped ion}\newline Well-isolated from environment & Physics package contained in volume of 0.54 m$^3$ & Ca & $2.3\times10^{-14}/\sqrt{\tau}$,\vspace{2mm} 10 to $3\times10^4$~s & $7.7\times10^{-17}$ &  Wuhan Inst., Chinese Academy of Sciences, Taizhou Univ.~\cite{wuhan1}  \\ 
			\hline
		\end{tabular}
		\label{tableoptical}
	\end{table*}

\subsection{Comparison of High-Performance Frequency Standards}
\begin{figure}
	\begin{center}
		\includegraphics[width=3.55in]{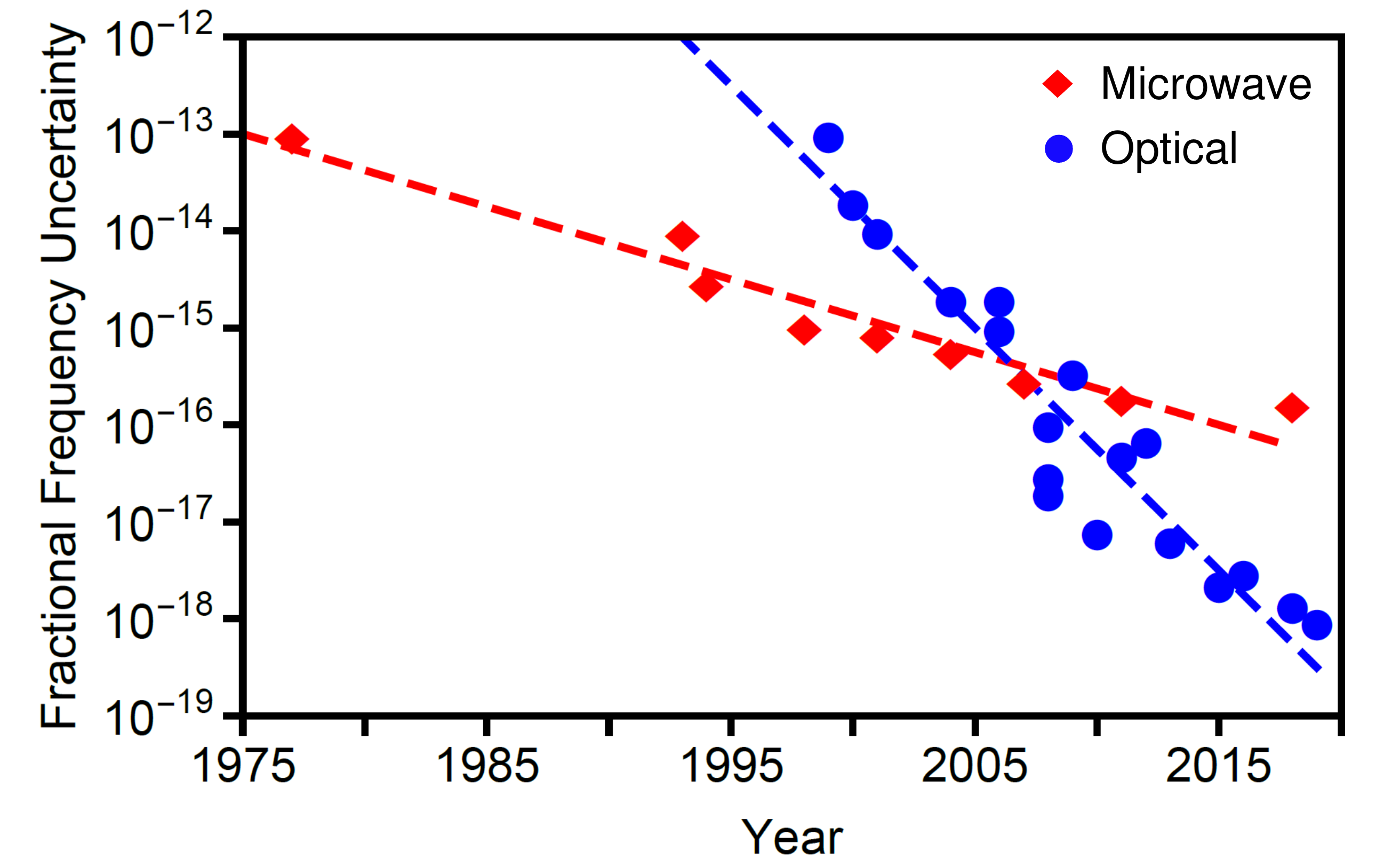}
		\caption{The evolution of the fractional frequency uncertainty of microwave and optical frequency standards reported in Refs.~\cite{Poli2013opticalrev} and~\cite{RevModPhys.90.025008} shown with additional data points from Refs.~\cite{Weyers_2018, McGrew2018, brewer}. The dashed curves are meant to guide the eye and emphasize the different trends in the state-of-the-art of the two types of frequency standards.}
		\label{clockuncertaintyhistory}
	\end{center}
\end{figure}

While it is still too early to define a best fit for the short-term instability vs.~size of optical frequency standards, it is reasonable to infer that such a trend will be distinct from that for microwave frequency standards. There has also been a distinct difference in the evolution of the fractional frequency uncertainty of microwave vs.~optical frequency standards over the past several decades, which has been analyzed in Refs.~\cite{Poli2013opticalrev} and~\cite{RevModPhys.90.025008} and reproduced here in Fig.~\ref{clockuncertaintyhistory} with the inclusion of additional data points from Refs.~\cite{Weyers_2018, McGrew2018, brewer}. These trends indicate that future improvements in microwave frequency standard performance will likely be incremental, whereas the pace of improvement of optical frequency standards will be more rapid.

Laboratory optical lattice frequency standards are likely to serve as the primary ground frequency standards in any future redefinition of the SI second. The instability and accuracy offered by these high-performance optical frequency standards will also enable new tests of fundamental physics including Einstein's theory of relativity~\cite{Chou1630}, which predicts that frequency standards will tick more slowly the closer they are to massive objects, and the variation of fundamental constants predicted by extensions of the Standard Model of particle physics~\cite{RevModPhys.90.045005}, which can be measured by comparing the frequencies of optical frequency standards over long time periods.

While the present work focuses on practical frequency standards, it is worthwhile noting the current status of record-holding instruments. In Table~\ref{highperformanceclocks}, we list recently reported instability and uncertainty measures for high-performance optical frequency standards and cesium fountains. Currently, optical frequency standards achieve approximately 1000x improved instability and 100x improved uncertainty compared to their microwave counterparts.

There are a number of efforts underway to improve the performance and reduce the SWaP of optical frequency standards. Drift of the laser between feedback cycles is the limiting factor in state-of-the art frequency standards~\cite{Poli2013opticalrev}, and a number of techniques are under investigation to improve the standalone laser system stabilization, including cavity cooling~\cite{Kessler2012}, vibration-insensitive cavity designs~\cite{PhysRevA.79.053829}, and using spectral holes in cryogenically cooled crystals as a reference~\cite{PhysRevLett.114.253902}. Novel methods to probe the atoms non-destructively~\cite{PhysRevA.79.061401} or to build frequency standards in tandem~\cite{PhysRevLett.111.170802} may also improve the frequency standard stability by lengthening accessible interrogation times without sacrificing SNR.

In addition to the efforts referenced above regarding miniaturization, there has also been substantial work towards improving the portability of frequency combs for optical frequency standards. The original titanium-sapphire laser-based combs have largely been replaced by fiber-based sytems~\cite{inaba2006long} or whispering gallery mode resonators~\cite{Kippenberg555}. More recently, there have also been a number of efforts towards miniaturizing microresonators and integrating them on chips, which represent promising engineering advances towards future practical optical frequency standards~\cite{Spencer2018, Bao2019, Newman:19}.

Recent demonstrations showed that a master optical frequency standard and a quartz-based microwave frequency standard separated by a 4 km free-space link could be synchronized via two-way time-frequency transfer with an ADEV at 1 second of $10^{-14}$~\cite{Bergeron:16}, using miniaturized frequency combs as the key component facilitating the link. A hybrid network linking optical frequency standards to microwave clocks could substantially alleviate reliance on GNSS for a broad range of applications including radar, navigation, and communications.

	\begin{table*}\centering
		\fontsize{0.31cm}{0.32cm}\selectfont
		\caption{Current high-performance frequency standards.}
		\label{highperformanceclocks}
		\renewcommand{\arraystretch}{1.4}
		\begin{tabular}{| >{\centering\arraybackslash}m{1.8cm} | >{\centering\arraybackslash}m{2cm} | >{\centering\arraybackslash}m{2.6cm} | >{\centering\arraybackslash}m{3.2cm} | >{\centering\arraybackslash}m{4.2cm} |}
			\hline
			\textbf{Frequency Standard Type}  & \textbf{Platform} & \textbf{Short-term instability} & \textbf{Uncertainty} & \textbf{Performer(s)}\\ \hline
			
			Optical & Sr Lattice & $4.8\times10^{-17}/\sqrt{\tau}$ & $2\times10^{-18}$ & JILA and NIST~\cite{YeRecentLattice} \\ \hline
			
			Optical & Yb Lattice & $1.5\times10^{-16}/\sqrt{\tau}$ & $1.4\times10^{-18}$ & NIST, UC Boulder, Peking Univ., Niels Bohr Institute, Istituto Nazionale di Ricerca Metrologica, Politecnico di Torino, Korea University~\cite{McGrew2018} \\ \hline
			
			Optical & Al$^+$ Quantum Logic & $1.2\times10^{-15}/\sqrt{\tau}$ & $9.4\times10^{-19}$ & NIST, UC Boulder, Univ. of Oregon~\cite{brewer}  \\ \hline
			
			Microwave & Cs Fountain & $2.5\times10^{-14}/\sqrt{\tau}$ & $1.71\times10^{-16}$ & PTB, Penn State Univ.~\cite{Weyers_2018} \\ \hline
			
			Microwave & Cs Fountain & $1.7\times10^{-13}/\sqrt{\tau}$ & ($1.1$~to~$4.4)\times10^{-16}$ & NIST, INRIM, Politecnico di Torino~\cite{Heavner_2014}  \\ 
			\hline
		\end{tabular}
	\end{table*}

\section{Conclusion}
In summary, we have presented an overview of the current state-of-the-art in the field of atomic frequency standards, including a summary of the physics of operation and performance for many products on the market today. The comparisons in Table~\ref{summary_table} and Fig.~\ref{TDEV_summary} provide the most complete summary and review of current products at the time of writing. Additionally, we describe the state-of-the-art of next-generation frequency standards in Section \ref{next_generation} and compare the projected performance of current and future products. We also discuss record performance at the time of writing for microwave and optical frequency standards.

In a broad sense, applications that require atomic frequency standards can be divided into three general areas: low-power applications, tactical applications, and strategic applications. Low-power applications are those that require extremely low power, on the order of less than 1~W, to achieve their mission; these applications currently require a CSAC. Tactical applications are those that have been serviced historically by the broad range of Rb gas-cell frequency standards on the market today, for a variety of communications and military applications. Strategic applications are those for which an investment in the high size, power, and cost of high-performance reference frequency standards is required to meet system performance.

Fueled by the demands of low-power commercial and military applications, we expect to see a significant change in the market availability of low-power atomic frequency standards within the next 5 to 10 years. There was only one commercially-available product for several years (Microchip CSAC), but the addition of a second vendor (Teledyne) as well as continued investment from Europe \cite{vicarini2018demonstration, karlen2019sealing}, China \cite{zhao2018new}, and Japan \cite{zhang2019ulpac} promise to continue to evolve the product landscape in the future. Similarly, we expect the high-performance product space (better than a Microchip 5071A Cesium Beam Tube) to undergo evolution as well. The recent introduction of cold atom commercial products by SpectraDynamics and Muquans, other commercial efforts by Microchip \cite{scherer2018analysis, tallant2019progress} and Honeywell \cite{7546775}, and research laboratory prototypes by JPL \cite{7437483}, NPL \cite{mulholland2019laser}, NIST \cite{Newman:19}, AFRL \cite{PhysRevApplied.9.014019}, and others \cite{commercialfountain} point to a more varied and competitive landscape for strategic frequency standards in the future. In comparison, we note that there is a comparative lack of commercial investment in novel mid-range tactical frequency standards, which suggests that these applications are well-served by the current range of Rb frequency standards on the market today. We expect continued changes in the low-power as well as strategic (microwave and optical) commercial atomic frequency standard landscape in the years to come.

\section*{Acknowledgments}
\noindent This work was funded by the MITRE Innovation Program. The authors thank John Betz and Erik Lundberg for a careful review of the manuscript.

Approved for Public Release; Distribution Unlimited. Public Release Case Number 20-0651. 


© 2020 IEEE.  Personal use of this material is permitted.  Permission from IEEE must be obtained for all other uses, in any current or future media, including reprinting/republishing this material for advertising or promotional purposes, creating new collective works, for resale or redistribution to servers or lists, or reuse of any copyrighted component of this work in other works.

\ifCLASSOPTIONcaptionsoff
  \newpage
\fi



%



\bibliographystyle{IEEEtran}
\bibliography{IEEEabrv,atomicclocks}

%

\begin{IEEEbiography}
{Bonnie L. Schmittberger} received the A.B. degree in physics from Bryn Mawr College in 2010 and the A.M. and Ph.D. degrees in physics from Duke University in 2013 and 2016, respectively. Her graduate research focused on nonlinear optics in ultracold atoms.

She is currently an Experimental Physicist at The MITRE Corporation in McLean, VA, where she is working on atomic and optical sensors for applications in communication and navigation. From 2016 to 2018 she was a postdoctoral researcher at the Joint Quantum Institute, where she worked with quantum states of light for applications in metrology. Dr. Schmittberger is a member of the Optical Society of America.
\end{IEEEbiography}

\begin{IEEEbiography}
{David R. Scherer} (M'12-SM'15) received the B.S. degree in \mbox{electrical} engineering from Boston \mbox{University} in 1999, and the M.S. and Ph.D. degrees in optical sciences from the University of Arizona in 2002 and 2007, respectively. His dissertation research focused on interferometry with ultracold atoms.

He is currently a Lead Scientist at The MITRE Corporation in Bedford, MA, where he is working on novel applications of atomic clocks. From 2012 to 2018, he was a Senior Physicist at Microchip, where he advanced several efforts related to next-generation atomic clocks, including development of a trapped-ion prototype clock as well as a demonstration of the world's smallest and lowest-power cold-atom system. From 2008 to 2012, he was a Principal Scientist at Physical Sciences Inc. (PSI), where he led several government-sponsored research projects in the areas of atomic sensors and integrated photonics.

Dr. Scherer is a member of ION and a Senior Member of IEEE and OSA. He currently serves as a committee member on IEEE Standard 1193, has served on the technical planning committee of the IEEE Frequency Control Symposium, and was the chair of the Boston Chapter of the IEEE Photonics Society. His research interests include applications of atomic clocks, precise timing, and quantum sensors for navigation, communication, and synchronization missions.

\end{IEEEbiography}







\end{document}